\documentclass[prc,showpacs,preprintnumbers,amsmath,amssymb,
superscriptaddress,floatfix,nofootinbib]{revtex4}

\usepackage{graphicx}
\usepackage{hyperref}
\hypersetup{hypertex=true,colorlinks=true,linkcolor=blue,
anchorcolor=blue,citecolor=blue}
\usepackage{dcolumn}
\usepackage{bm}
\usepackage[mathlines]{lineno}
\usepackage{epsfig}  
\usepackage{color}
\usepackage{slashed}
\usepackage{float}
\usepackage{xcolor}
\usepackage{subfigure}
\usepackage{cancel}
\usepackage{comment}
\usepackage{placeins}

\makeatletter
\AtBeginDocument{%
  \expandafter\renewcommand\expandafter\subsection\expandafter
    {\expandafter\@fb@secFB\subsection}%
  \newcommand\@fb@secFB{\FloatBarrier
    \gdef\@fb@afterHHook{\@fb@topbarrier \gdef\@fb@afterHHook{}}}%
  \g@addto@macro\@afterheading{\@fb@afterHHook}%
  \gdef\@fb@afterHHook{}%
}
\makeatother

\begin{document}

\title{Resolving the W boson Mass in the Lepton Specific Two Higgs Doublet Model}
\author{Ali Çiçi}
\email{ali.cici@cern.ch}
\affiliation{Republic of Turkey National Education Ministry}
\author{Hüseyin Dağ}
\email{huseyin.dag@cern.ch}
\affiliation{Bursa Technical University}

\begin{abstract}

In 2022, the CDF Collaboration reported the $W$-boson mass, $M_W=80.4335\pm0.0094~\mathrm{GeV}$, which deviates from the Standard Model (SM) prediction, $M_W^{\rm SM}=80.357\pm0.006~\mathrm{GeV}$, by about $7\sigma$. By contrast, the CMS Collaboration obtained $M_W=80.3602\pm0.0099~\mathrm{GeV}$, very close to the SM global electroweak fit value of $\sim80.357~\mathrm{GeV}$. Motivated by this situation, we reassess the $W$-boson mass within the Lepton-Specific Two Higgs Doublet Model (LS-2HDM). We perform random scans (generated with SARAH 4.13.0 and evaluated with SPheno 4.0.3) and confront the results with up-to-date theoretical and experimental constraints. The scan enforces vacuum stability, perturbative unitarity, and perturbativity; electroweak precision observables via the oblique parameters $(S,T,U)$; LEP bounds on $H^\pm$; rare $B$-meson decays; lepton flavor universality (LFU) in $Z$ and $\tau$ decays; and LHC 13 TeV searches for additional Higgs bosons. Viable points are further tested with HiggsTools (HiggsSignals + HiggsBounds). In the LS-2HDM, if $h_1$ is the SM-like Higgs at $m_{h_1}\simeq125$ GeV with $|\cos(\beta-\alpha)|\lesssim0.06$, $17\lesssim\tan\beta\lesssim39$, $144\lesssim m_{h_2}\lesssim414$ GeV, and $435\lesssim m_{A,H^{\pm}}\lesssim685$ GeV, the model reproduces the 2024 CMS $W$-boson mass within $3\sigma$. Solutions near the 2022 CDF value, $M_W=80.4335\pm0.0094~\mathrm{GeV}$, survive; however, after applying all constraints, including HiggsTools, they approach it at best within $\lesssim2\sigma$. Our findings emphasize that the LS-2HDM favors the CMS results consistently with the current experimental results. On the other hand, while one can accommodate also the CDF results in this model theoretically, up-to-date electroweak precision bounds on the oblique parameters $(S,T,U)$ together with the SM-like Higgs and LFU constraints  exclude these solutions and our results for $W-$boson mass can be only as close as about $2\sigma$ to the CDF results.

Keywords: $W-$boson, Lepton Specific Two Higgs Doublet Model, CDF, CMS

\end{abstract}
\maketitle

\section{Introduction} 

The Standard Model (SM) has withstood rigorous testing, successfully been explaining various phenomena in particle physics. However, recent measurements of the $W-$boson mass at the Collider Detector at Fermilab (CDF) have revealed substantial discrepancies between experimental observations and theoretical predictions within the SM framework. The CDF reported a precise measurement of the $W-$boson mass using the $8.8 fb^{-1}$ dataset  from $p\bar{p}$ collisions with a center-of-mass energy of 1.96 TeV as \cite{CDF:2022hxs}

\begin{align}
    M_W^{\rm CDF} = 80.4335 \pm 0.0094 \ \text{GeV},
\label{eq:mwcdf}
\end{align}
which deviates from the SM prediction  $M_W^{\rm SM} = 80.357 \pm 0.006 {\rm GeV}$ by $7\sigma$ \cite{Blum:2013xva,RBC:2018dos,Keshavarzi:2018mgv,Davier:2019can,Aoyama:2020ynm,Colangelo:2018mtw,Hoferichter:2019mqg,Melnikov:2003xd,Hoferichter:2018kwz,Blum:2019ugy,ParticleDataGroup:2020ssz,ATLAS:2017rzl}. Such a notable discrepancy suggests the potential existence of new physics phenomena, necessitating a comprehensive exploration of physics Beyond the Standard Model (BSM).

On the other hand, the $W-$boson mass result from the CMS experiment in 2024 — following almost a decade of research — has eased considerably the strain in the literature created by the CDF measurement. With the value of $M_W = 80.3602 \pm 0.0099 \text{GeV}$, the CMS Collaboration obtained a value very close to the Standard Model's global electroweak fit prediction of around 80.357 GeV \cite{CMS:2024lrd}. This new measurement largely resolves the long-standing tension displayed by CDF, moderating the anomaly that first put the Standard Model at risk. However, the values of CDF and CMS disagree with each other: their central values differ by approximately 73 MeV (some $5\sigma$). Therefore, when these two experimental results are considered together, a more conservative view about the need for new physics is appropriate.

The inclusion of BSM particles might induce quantum corrections accountable for the deviation observed in the $W-$boson mass, hence they have sparked a growing interest in exploring new physics BSM, mostly by altering the oblique parameters, $S$, $T$, and $U$. These approaches include a broad range of theoretical frameworks, including effective field theory methods \cite{deBlas:2022hdk,DiLuzio:2022xns,Paul:2022dds,Balkin:2022glu,Endo:2022kiw,Cirigliano:2022qdm,Gupta:2022lrt}, supersymmetric (SUSY) models \cite{Du:2022pbp,Yang:2022gvz,Athron:2022isz,DiLuzio:2022ziu,Zheng:2022irz,Ghoshal:2022vzo,Peli:2022ybi} , leptoquark models \cite{Athron:2022qpo,Cheung:2022zsb,Bhaskar:2022vgk}, gravitational approaches \cite{Liu:2022jdq,Addazi:2022fbj}, Little Higgs models \cite{Strumia:2022qkt}, and extensions of the SM involving additional scalar singlets \cite{Sakurai:2022hwh,Cheng:2022aau,Faraggi:2022emm,Cai:2022cti} or triplets \cite{Asadi:2022xiy,FileviezPerez:2022lxp,Kanemura:2022ahw}. Additionally, models featuring vector-like leptons have been explored \cite{Lee:2022nqz,Gu:2022htv,Crivellin:2022fdf,Kawamura:2022uft,Nagao:2022oin,Popov:2022ldh}, alongside investigations from the point of view of neutrino masses and seesaw mechanisms \cite{Cheng:2022jyi,Borah:2022obi}, and other therotical approaches 
\cite{Du:2022brr,Mondal:2022xdy,Carpenter:2022oyg,Zeng:2022lkk,Du:2022fqv,Borah:2022zim,Chen:2022ocr,TranTan:2022kpq,Batra:2022pej,Zhang:2022nnh,Heckman:2022the,Bandyopadhyay:2022bgx,Senjanovic:2022zwy}. Furthermore, Two Higgs Doublet Models (2HDM) have been attracting a considerable interest by providing a simple extension of SM \cite{Zhu:2022tpr,Lu:2022bgw,Song:2022xts,Bahl:2022xzi,Babu:2022pdn,Biekotter:2022abc,Heo:2022dey,Ahn:2022xax,Arcadi:2022dmt,Chowdhury:2022moc,Lee:2022gyf,Abouabid:2022lpg,Benbrik:2022dja,Kim:2022xuo,Hashemi:2023tej,Azevedo:2023zkg,BellinatoGiacomelli:2023tnj,Chakrabarty:2022dai,Atkinson:2022qnl,Wang:2022yhm,Ghorbani:2022vtv}, while the Higgs bosons in its spectrum can interfere with the SM particles they can yield some deviations in $W-$boson mass, Lepton Flavor Universality (LFU) \cite{Angel:2013hla,Chun:2016hzs,Wang:2018hnw} , muon g-2 \cite{Davier:2010nc,Hagiwara:2011af,Borsanyi:2020mff} through their radiative contributions. Among these efforts, certain studies specifically target the $W-$boson mass discrepancy while also addressing the dark matter problem by choosing a focused approach on both issues within their framework \cite{Zhu:2022tpr,Strumia:2022qkt,Sakurai:2022hwh,Lee:2022nqz,Kawamura:2022uft,Zhang:2022nnh,Zeng:2022lkk,Borah:2022zim}. It appears that this anomaly attracts a lot of attention from several theoretical approaches, among which 2HDMs are an important class. However, while 2HDMs are widely used in understanding the $W-$boson mass anomaly, their parameter space is limited by experimental data from several experiments \cite{Angel:2013hla,Chun:2016hzs,Abe:2015oca,Wang:2018hnw,HFLAV:2022esi,ATLAS:2020zms,ATLAS:2021upq,ATLAS:2017jag,ATLAS:2017otj,CMS:2023fod}. Therefore, in this study, motivated by these theoretical and experimental efforts within 2HDM frameworks, an exploratory investigation aimed at reconciling the $W-$boson mass discrepancy by scanning possible solutions within the 2HDM parameter space.

This study focuses on investigating the impact of the parameter space of 2HDM on the $W-$boson mass. Two crucial considerations are taken into account in our analyses: theoretical limitations and compatibility with the current experimental data. Theoretical limitations arise from constraints related to stability of the scalar potential and perturbativity. The predictions of 2HDMs must also align with the outcomes of various experiments, such as those involving rare decays of B-meson, $Z-$boson decay, tau-lepton decay, and observations from the Large Hadron Collider (LHC). To explore the implications of different types of 2HDMs within these theoretical and experimental limitations, the HiggsTools framework was utilized  \cite{Bahl:2022igd}. 
Among several types of 2HDMs, some models can have stronger motivation by distinguishing the SM fermions based on their assigned $Z_{2}$ symmetries. For instance, the Lepton Specific 2HDM (LS-2HDM) assigns a $Z_{2}$ symmetry for the SM fermions such that the quarks interact with one Higgs doublet, while the leptons with another doublet. In this way, one can impose a mass hierarchy between the quarks and leptons. In addition, this discrimination between the quarks and leptons yield different results in production of the heavy Higgs bosons in the collider experiments \cite{CMS:2023fod}, and hence, the current strong limits from the recent CMS findings on scalar masses of the 2HDM can be modified considerably. For instance, the production of the extra Higgs bosons in the pp collisions within the LS-2HDM framework is suppressed with $\tan\beta$ parameter, and they can escape from the detection, while they contribute to the gauge boson masses and LFU processes at the loop level significantly. Despite its different behaviour there are still possible experimental tested and the limits for the models such as those in the LS-2HDM class \cite{CMS:2019qcx}.

The following steps are followed in this work. The scalar sector mass spectra that are consistent with both theoretical and experimental requirements are first obtained. After determining these solutions, an exploration of implications on Lepton Flavor Universality is carried out by considering the processes involving the tau-lepton and $Z-$boson. These analyses further constrain the parameter space of the LS-2HDM. Then, we examine the consistency of the LS-2HDM with both the CDF and CMS $W$-boson mass measurements under all theoretical and experimental constraints, and we show that once the electroweak oblique parameter bounds $(S, T, U)$ and SM-like Higgs boson requirement  are imposed, all solutions within about 1$\sigma$ deviation of the CDF W-boson mass are excluded. The closest viable points lie at $\lesssim 2\sigma$, whereas the CMS value is readily accommodated.

Further, this comprehensive investigation of the LS-2HDM discusses the implications of both recent measurements. The motivation of our study is to explain the potential excess reported by CDF in the $W-$boson mass in terms of the LS-2HDM parameter space, and also verify whether such solutions are still in agreement with the CMS measurement. To this end, the LS-2HDM parameter space is scanned systematically, incorporating theoretical consistency conditions (such as scalar potential stability and perturbativity) and available experimental constraints (such as rare B-meson decays, $Z-$boson and tau-lepton data, LHC searches, etc.). Subsequently, parameter space regions that can explain the CDF and CMS experiment W-mass measurements are evaluated and the model consistency with each result is commented on. In the following sections, the stabilizing effect of the CMS measurement on LS-2HDM literature will be emphasized, and it will be shown that our findings are in agreement with both experimental results.

This work is structured as follows: The second section summarizes LS-2HDM, focusing on its Higgs sector and Yukawa interactions. The third section outlines the theoretical and experimental constraints used in this analysis, along with a detailed discussion of their impacts on the parameter space. In the fourth section, the parameter space of the LS-2HDM is systematically explored, and potential solutions are provided. Finally, the last section offers discussions and concluding remarks.

\section{Lepton Specific 2HDM}

2HDMs are obtained by extending the scalar sector of the SM with the addition of a second scalar doublet possessing the same quantum numbers as the SM Higgs Doublet. The gauge group of the 2HDMs is identical to that of the SM. In 2HDMs, eight scalar fields are introduced and following a spontaneous symmetry breaking process akin to that of the SM, three of these fields confer masses of the gauge bosons, while the remaining five fields undergo mixing, resulting in five distinct physical scalar bosons. These additional degrees of freedom present in the 2HDM have far-reaching implications for Higgs boson phenomenology. The scalar doublets of 2HDMs are given as 

\begin{equation}
\Phi_{1}=\left(\begin{array}{c}
\phi_{1} + i\phi_2 \\
\phi_{3} + i\phi_4
\end{array}\right) \qquad\text{and} \qquad \Phi_{2}=\left(\begin{array}{c}
\phi_{5} + i\phi_6 \\
\phi_7 + i\phi_8
\end{array}\right),
\label{sekizalan}
\end{equation}
where $\phi_3$ and $\phi_7$ develop non-zero vacuum expectation values (VEVs) as $<\phi_3> = v_1/\sqrt{2}$ and $<\phi_7> = v_2/\sqrt{2}$ satistfying $v_{SM}=\sqrt{v_1^2+v_2^2}$. The particle content of 2HDMs in the scalar sector are two CP-even ($h_{1,2}$), one CP-odd ($A$), and two charged ($H^\pm$) Higgs bosons. For a comprehensive review of 2HDMs, please refer to Ref. \cite{Diaz:2002tp,Bhattacharyya:2013rya,Das:2015qva,Das:2015mwa,Das:2018qjb} and the citations within there.

In 2HDMs, the addition of a second scalar doublet in the Yukawa sector leads to Yukawa couplings that lack of flavor-diagonal properties, thereby resulting in the emergence of tree-level FCNC processes which are severely constrained from experiments. To resolve this issue, a viable approach is to introduce a discrete symmetry to the scalar and Yukawa potentials \cite{Ko:2012hd}. The application of this symmetry limits the interactions between the additional scalar doublet and fermions, suppresses flavor-changing neutral currents at tree-level, and establishes the model as a viable approach for satisfying experimental constraints. One important example of this discrete symmetry is the $Z_{2}$ symmetry \cite{Ko:2012hd,Hashemi:2018kct,Han:2018znu,Nomura:2019wlo}, which is described as 

\begin{equation}
\begin{aligned}
&\Phi _{1} \rightarrow -\Phi _{1}\;\;\text{and\ \ }\Phi _{2}\rightarrow
\Phi _{2}~, \\
&D_{j} \rightarrow  D_{j},\;\;U_{j}\rightarrow U_{j}\;\;\text{and \ \ }E_{j} \rightarrow - E_{j}~,
\end{aligned}
\label{simetriLS}
\end{equation}
where $E_j$ denotes right-handed leptons, $U_j$ and $D_j$ denote right-handed up-type and down-type quarks. The 2HDMs with this choice of $Z_2$ symmetry are often called as Lepton Specific 2HDM (LS-2HDM). Under $Z_2$ symmetry, the tree level potential of the LS-2HDM becomes

\begin{equation}
\begin{aligned}
V_{\rm tree} = & m_{1}^2|\Phi_1|^2+m_{2}^2|\Phi_2|^2 - \left(m_{3}^2\Phi_1^{\dagger}\Phi_2+h.c.\right)
  +\frac{\lambda_{1}}{2}|\Phi_1|^4 + \frac{\lambda_{2}}{2}|\Phi_2|^4 \\     &+\lambda_{3}|\Phi_1|^2|\Phi_2|^2 +\lambda_{4}|\Phi_1^{\dagger}\Phi_2|^2 
  +\frac{\lambda_5}{2}\left[(\Phi_1^\dagger \Phi_2)^2+h.c.\right],
\end{aligned}
\label{eq:skaler}
\end{equation}
where $m_{1,2}$ correspond to the mass terms of the scalar potentials, and $\lambda_{1,..,5}$ refer to self-couplings. In Eqn. \ref{eq:skaler}, the term involving $m_3^2$ arises from a combination of scalar doublets and violates the $Z_2$ symmetry, resulting in soft symmetry breaking. The expressions for the masses of extra scalar bosons at tree level are obtained from Eqn. \ref{eq:skaler} as

\begin{equation}
\begin{aligned}
& m_{H^\pm}^2 =\frac{1}{2}\left(m_3^2\frac{1}{\sin\beta\cos\beta} -\lambda_3 v_{\text{SM}}^2  \right),  \\
&m_{h_{1,2}}^2=\frac{1}{4}m_3^2 \left(\tan\beta + \cot\beta \mp \frac{2}{\sin 2\alpha} \right)  + \lambda_1 \cos^2\beta v_{\text{SM}}^2 + \lambda_2 \sin^2\beta v_{\text{SM}}^2 \mp\frac{\lambda_3 \sin\beta \cos\beta v_{\text{SM}}^2}{\sin 2\alpha} \mp\frac{\lambda_5 \sin\beta \cos\beta v_{\text{SM}}^2}{\sin 2\alpha}, \\
&m_{A}^2=\frac{1}{2}\left(m_3^2\frac{1}{\sin\beta\cos\beta} -\lambda_3 v_{\text{SM}}^2 + \lambda_4 v_{\text{SM}}^2 \right).
\end{aligned}
\label{extraHiggsMasses}
\end{equation}
where $\sin\alpha$ is the mixing angle of CP-even Higgs bosons and $\tan\beta = v_2/v_1$ is the ratio of VEVs of doublets. In addition to these expressions in tree-level, radiative corrections to scalar masses must be taken into account. These corrections can be calculated through one-loop improved scalar potential described as \cite{Ferreira:2015pfi,Cici:2019zir}

\begin{align}
   V=V_{\rm tree} + V_{\rm loop},
\end{align}
where the loop potential is described as 
\begin{equation}
V_{\rm loop} = \dfrac{1}{64\pi^{2}}\displaystyle \sum_{\alpha}n_{\alpha}m_{\alpha}^{4}\left[\log\left(\dfrac{m_{\alpha}^{2}}{\mu^{2}} \right) - \dfrac{3}{2}\right].
\label{eq:loopV}
\end{equation}
with 
\begin{equation}
n_{\alpha} = (-1)^{2s_{\alpha}}Q_{\alpha}C_{\alpha}(2s_{\alpha}+1),
\end{equation}
where $\mu$ is the renormalization scale, $m_{\alpha}$ denote the masses of particles contributing at loop-level, $s_{\alpha}$ are the spin of the particles, $Q_{\alpha}=1(2)$ for neutral (charged) particles and $C_{\alpha}=3 (1)$ for quarks (leptons), and $\alpha$ runs over all the particles that couple to the scalars at tree-level. Furthermore, The Yukawa Lagrangian under $Z_2$ symmetry can be written as \cite{Cici:2019zir}
\begin{equation}
\mathcal{L}_{\mathit{Y}}= -Y_e^{ij}\bar{L}_i\Phi_1 E_j - Y_u^{ij} \bar{Q}_i\Phi_2^C U_j 
- Y_d^{ij} \bar{Q}_i\Phi_2 D_j+h.c. ~, 
\label{intro_eq:yukawa}
\end{equation}
where $Y_{e,u,d}^{ij}$ are the Yukawa couplings, and $L_i$ and $Q_i$ denote the $SU(2)_L$ doublets for leptons and quarks, respectively. Similar to the Yukawa couplings in the SM, the effective Yukawa couplings for the LS-2HDM are obtained as shown in Table \ref{tab:yuk}. Since the mass of the top quark is primarily determined by $\Phi_2$, the value of $\tan\beta$ is constrained as $v_2 > v_1$. If $v_2 \gg v_1$ and $\tan\beta \gg 1$, the Yukawa coupling of the CP-odd Higgs boson with quarks becomes negligibly small, while its coupling with leptons increases.

\begin{table}[hbt!]
    \centering
\renewcommand{\arraystretch}{1.5} 
\setlength{\tabcolsep}{16pt}
    \begin{tabular}{|c|c|c|c|}
\hline
$Y_u^{h_1},Y_d^{h_1},Y_l^{h_1} $& $\cos\alpha /\sin\beta$ & $\cos\alpha /\sin\beta$  & $-\sin\alpha/\cos\beta$   \\
\hline
$Y_u^{h_2},Y_d^{h_2},Y_l^{h_2} $& $\sin\alpha /\sin\beta$ & $\sin\alpha/\sin\beta$  & $\cos\alpha /\cos\beta$   \\
\hline
$Y_u^{A},Y_d^{A},Y_l^{A} $& $\cot\beta$ & $-\cot\beta$  & $\tan\beta$   \\
\hline
    \end{tabular}
    \caption{Effective Yukawa couplings for LS-2HDM}
    \label{tab:yuk}
\end{table}

\section{Theoretical and Experimental Constraints}
\label{teorikvedeneyselsinir}

In this section, theoretical and experimental constraints on the parameter space of the LS-2HDM are investigated to determine the allowed parameter regions for our analysis. First, theroretical limitations on the scalar potential of the LS-2HDM are considered. The scalar potential given in \ref{eq:skaler} and \ref{eq:loopV} should adhere to the vacuum stability and perturbativity conditions which limits the self couplings as \cite{Cao:2009as}

\begin{equation}
\lambda_i < 4\pi ~ (i=1,\ldots,5).
\label{eq:pert} 
\end{equation}
In order to provide unitarity at tree level, self couplings must satisfy the relations \cite{Cao:2009as} given by

\begin{equation}
\begin{aligned}
&3(\lambda_1+\lambda_2)\pm\sqrt{9(\lambda_1-\lambda_2)^2+4(2\lambda_3+\lambda_4|)^2}
<16 \pi,  \\
&\lambda_1+\lambda_2\pm\sqrt{(\lambda_1-\lambda_2)^2+4|\lambda_5|^2}
<16\pi, \\
&\lambda_1+\lambda_2\pm\sqrt{(\lambda_1-\lambda_2)^2+4|\lambda_5|^2} <16 \pi, \\
&\lambda_3+2\lambda_4\pm 3 | \lambda_5|<8\pi,\\
&\lambda_3\pm\lambda_4 < 8\pi,\\
& \lambda_3\pm|\lambda_5|<8\pi.
\end{aligned}
\label{eq:PertBounds}
\end{equation}
Furthermore to ensure that the scalar potential of the LS-2HDM is finite, free of flat directions, and stable at large field values, following conditions

\begin{equation}
\begin{aligned}
&  \lambda_{1,2}>0,  \\
&  \lambda_3>- \sqrt{\lambda_1\lambda_2},  \\
&  \lambda_3+\lambda_4-|\lambda_5|> - \sqrt{\lambda_1\lambda_2},
\end{aligned}
\label{eq:VacStabBounds}
\end{equation}
are imposed.

Besides the aforementioned theoretical constraints, the LS-2HDM is subject to stringent experimental constraints, as well. It is evident that the predictions of LS-2HDM should align with a wide range of experimental observations, including precision measurements of the electroweak sector and collider searches for new particles and phenomena. First, the constraint on the charged scalar boson mass was determined from the Large Electron-Positron Collider (LEP) data as $m_{H^\pm} > 80$ GeV \cite{LEPHiggsWorkingGroupforHiggsbosonsearches:2001ogs}. As discussed in the previous section, due to the significant influence of $m_{h_1}$ on $m_t$, ${h_1}$ is identified as the SM-like Higgs boson, and its mass is fixed at $m_{h_1} \simeq 125 \pm 2$ GeV \cite{ATLAS:2012yve,CMS:2012qbp}, since the theoretical calculations of the Higgs boson mass involve about 2 GeV uncertainty \cite{AdeelAjaib:2013dnf,Gogoladze:2014hca}. Moreover, experimental data from electroweak precision measurements to the $W-$boson mass in BSM models are used to determine the oblique parameters $S$, $T$ and $U$, and they are constrained as  \cite{ParticleDataGroup:2024cfk},

\begin{equation}
\begin{aligned}
&S = -0.04 \pm 0.10, \\
& T = 0.01 \pm 0.12, \\
& U = -0.01 \pm 0.09, 
\end{aligned}
\label{eq:oblparam}
\end{equation}
where the expressions of these parameters will be given in Section \ref{SectionD}.

In addition, limitations from rare B-meson decays, such as $B_s \rightarrow \mu^+ \mu^-$ and $B_s \rightarrow X_s \gamma$ should be considered, since they are sensitive probes for extra scalars of BSM. In the LS-2HDM framework, these decay processes receive contributions from scalar states via loops, which limits models parameter space. Therefore, the following bounds on the rare B-meson decay branching ratios \cite{CMS:2020rox,Belle-II:2022hys,HFLAV:2022esi}

\begin{equation}
\begin{aligned}
& 1.95\times 10^{-9} \leq BR(B_s \rightarrow \mu^+ \mu^-) \leq 3.43 \times 10^{-9} \; (2\sigma) \\
& 2.99 \times 10^{-4} \leq BR(B_s \rightarrow X_s \gamma) \leq 3.87 \times 10^{-4} \; (2\sigma)
\end{aligned}
\label{eq:rareBdecay}
\end{equation}
are applied to the parameter space. Finally, parameters of LS-2HDM which estimate $m_W$ within 3$\sigma$ vicinity of the CDF result given in Eqn. (\ref{eq:mwcdf}) are considered in this work. 

Following these considerations, constraints on the parameters of LS-2HDM can be summarized in the following four groups. 

\begin{description}
\item[G1] Theoretical constraints:
\begin{itemize}
\item The self couplings should satisfy the ranges given in Eqns. \ref{eq:pert}, \ref{eq:PertBounds}, and \ref{eq:VacStabBounds}.
\end{itemize}
\item[G2] Experimental constraints:
\begin{itemize}
\item $m_{H^\pm} \geq 80$ GeV,
\item $-0.14 \leq S \leq 0.06$,  $-0.11 \leq T \leq 0.13$, $-0.1 \leq U \leq 0.08$,  $(1\sigma)$,
\item $1.95\times 10^{-9} \leq BR(B_s \rightarrow \mu^+ \mu^-) \leq 3.43 \times 10^{-9} (2\sigma)$,
\item $ 2.99 \times 10^{-4} \leq BR(B_s \rightarrow X_s \gamma) \leq 3.87 \times 10^{-4}(2\sigma) $.
\end{itemize}
\item[G3] $h_1$ is chosen to be SM-like Higgs boson with
\begin{itemize}
\item $m_{h_1} = 125.0 \pm 2$ GeV.
\end{itemize}
\item[G4] Constraint on $W-$boson mass reads
\begin{itemize}
\item  $80.4053 \leq  M_W^{\rm LS-2HDM} \leq 80.4617$ GeV ($3\sigma$ CDF).
\end{itemize}
\item[G5] Constraint on $W-$boson mass reads
\begin{itemize}
\item  $80.3305 \leq  M_W^{\rm LS-2HDM} \leq 80.3899$ GeV ($3\sigma$ CMS).
\end{itemize}
\end{description}
Labeling of constraints are chosen for brevity in further discussions. Note that the constraint on the Higgs boson in G3 is applied only on its mass in the first step of our analyses. We employ these constraints only to explore the regions where the deviation in $M_{W}$ can be realized within the allowed ranges reported by the CDF and CMS collaborations. However, in the second step, we perform further analyses and we also employ HiggsTools (HiggsSignals and HiggsBounds) \cite{Bahl:2022igd} to ensure the consistency of the Higgs boson solutions beyond its mass.

\section{EXPLORING THE PARAMETER SPACE}
\label{put}

In this section, the parameter space of the LS-2HDM is explored by performing a random scan of potential parameters using SPheno 4.0.3, generated via SARAH 4.13.0 \cite{Staub:2013tta,Porod:2011nf}. In these scans, solutions which satisfy the electroweak symmetry breaking condition, $(v_1^2 + v_2^2 \simeq v_{\rm SM}^2)$ are accepted. To ensure that the results of our random parameter scans are consistent with current measurements of the $W-$boson mass, the range of the self couplings are chosen as

\begin{equation}
\begin{aligned}
0 &\leq \lambda_{1,2} \leq \pi, \\
0 &\leq \lambda_3 \leq 4\pi,  \\
-2\pi &\leq \lambda_{4,5} \leq 0,  \\
-5&\leq m_3^2 \leq 5 \; {\rm TeV}^2, \\
1.2 &\leq \tan\beta \leq 40.0 \; . 
\end{aligned}
\label{eq:ranges}
\end{equation}

In determining the ranges of the parameters, we stay in the intervals allowed by the perturbativity. The constraints from perturbativity are applied staightforwardly to the couplings by following the condition given in Eqn.(\ref{eq:pert}). On the other hand, it is not straight forward for the $\tan\beta$ parameter. To adjust the range for this parameter we consider the couplings of the Higgs bosons to the SM fermions. To keep the top quark Yukawa coupling perturbative at all energy levels from $M_{Z}$ to some high energies, $\tan\beta$ should be bounded from below at about 0.3 \cite{Branco:2011iw}. In our scans, we lift this lower bound to 1.2 to ensure the fields in $\Phi_{2}$ form the SM-like Higgs boson. Similarly, one can also put an upper bound by following the perturbativity of the gauge couplings as well as the Yukawa couplings, which disfavors the solutions with $\tan\beta \gtrsim 30$ \cite{Akeroyd:2000wc,Arhrib:2000is}. These bounds, on the other hand, have been obtained in a general manner. As mentioned before, the behaviour of the Higgs bosons may differ in LS-2HDM from the other types of 2HDMs, so we put an upper limit on $\tan\beta$ at $40$. Apart from these constraints, we restrict the parameters further for practical reasons to optimize our scans to explore CDF and CMS compatible $M_{W}$ solutions. After successively applying the constraints listed in the previous section, solutions and their respective color coding in plots, which will be used throughout the rest of this work.
\subsection{Mass Spectrum of LS-2HDM}

\begin{figure}[t!]
\centering
\subfigure{\includegraphics[scale=0.35]{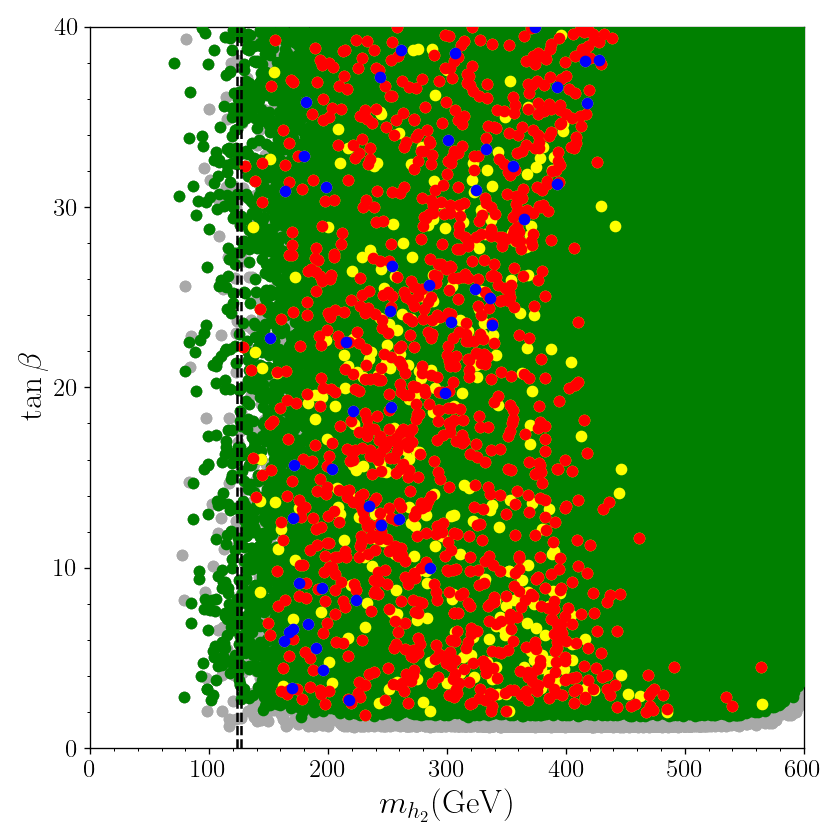}}%
\subfigure{\includegraphics[scale=0.35]{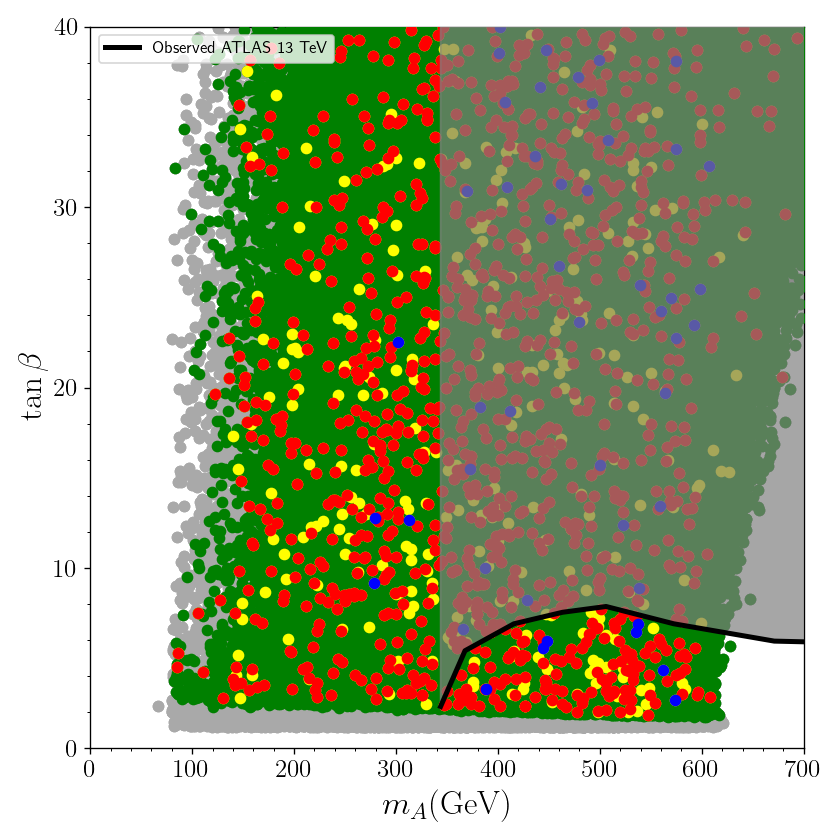}}\\%
\subfigure{\includegraphics[scale=0.35]{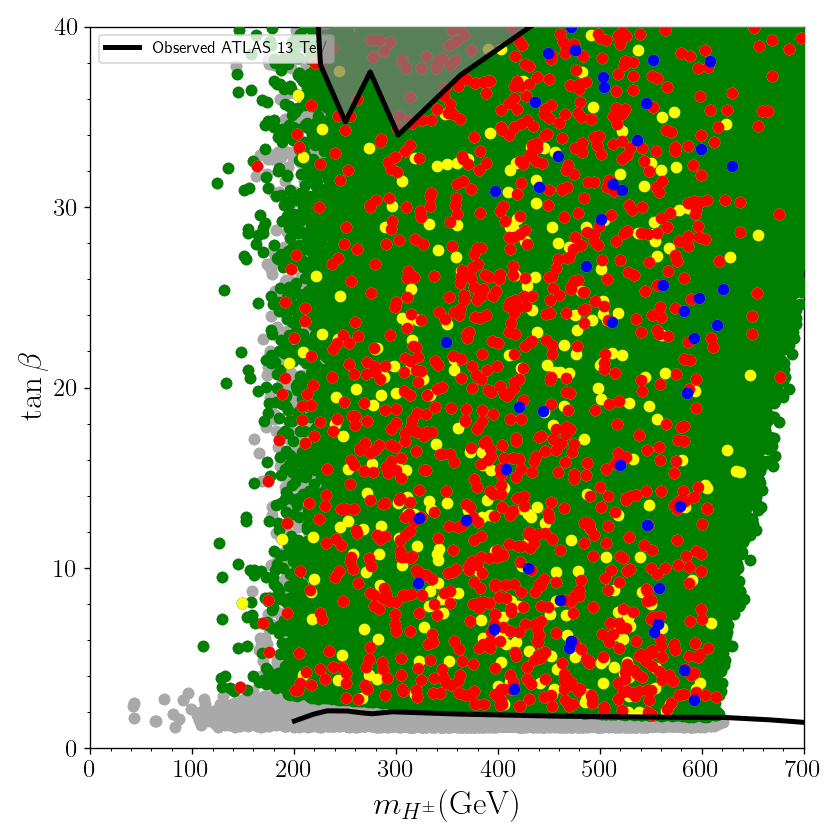}}
\subfigure{\includegraphics[scale=0.35]{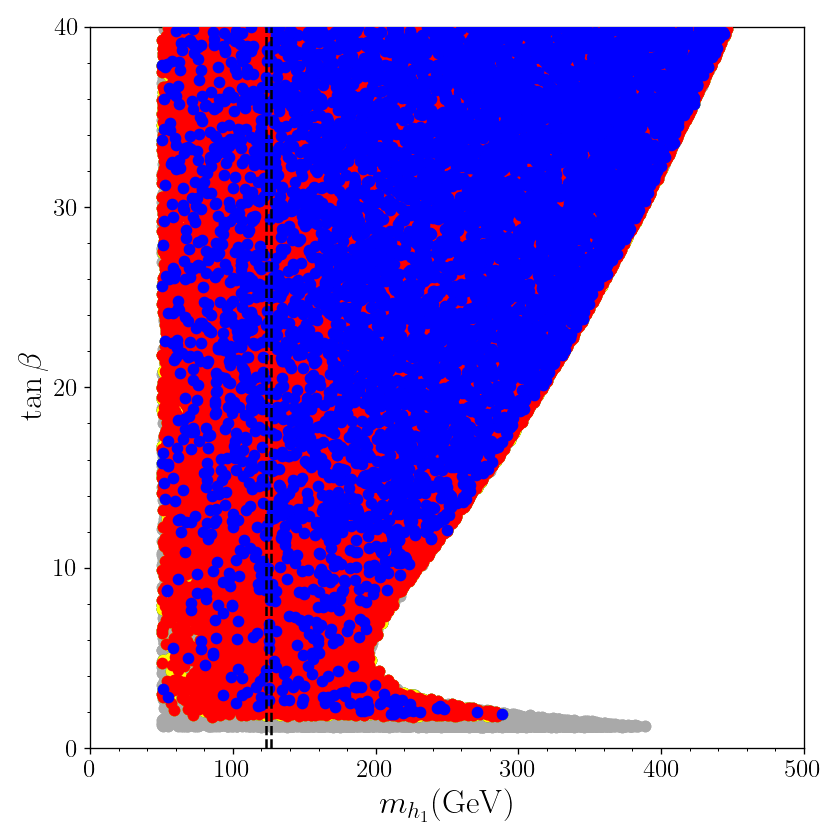}}%
\caption{ The relation between $\tan\beta$ and the mass of the LS-2HDM scalars, for $m_{h_2}$ (top-left), $m_A$ (top-right), $m_{H^\pm}$ (bottom-left) and $m_{h_1}$ (bottom-right). Color coding of solutions are described in detail in previous text and summarized as: Green points satisfy conditions G1 and G2, yellow points satisfy conditions G1, G2 and G3, blue points satisfy conditions G1, G2, G3 and G4, red points satisfy conditions G1, G2, G3 and G5. Solid black lines indicate ATLAS 13 TeV analysis observation limits and gray shaded areas are excluded by these analysis \cite{ATLAS:2020zms,ATLAS:2021upq}. In $\tan\beta$ vs $m_{h_1}$ plot (bottom-right), condition G3 is not applied. Dashed vertical lines represent the SM-like Higgs mass.}
\label{fig:Hmasses_1}
\end{figure}

\begin{figure}[t!]
\centering
\subfigure{\includegraphics[scale=0.35]{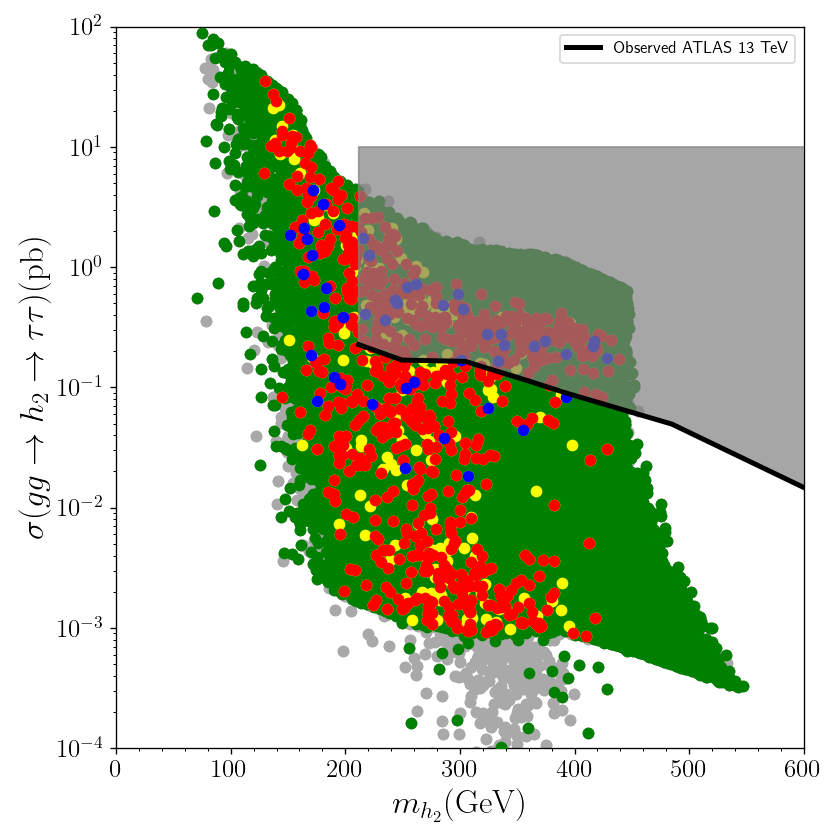}}
\subfigure{\includegraphics[scale=0.35]{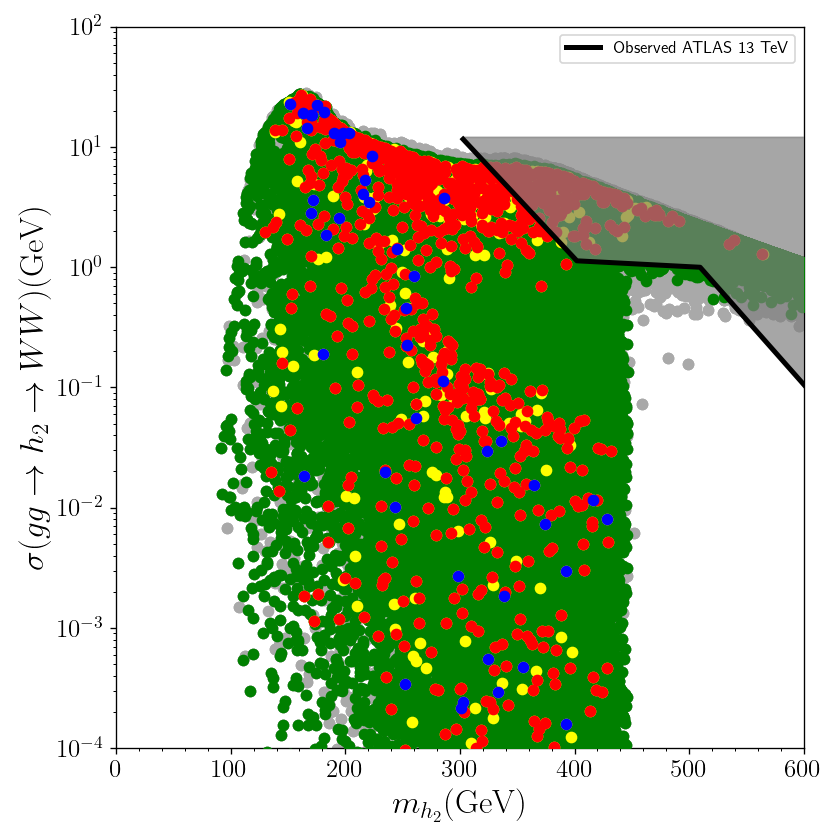}}\\%
\subfigure{\includegraphics[scale=0.35]{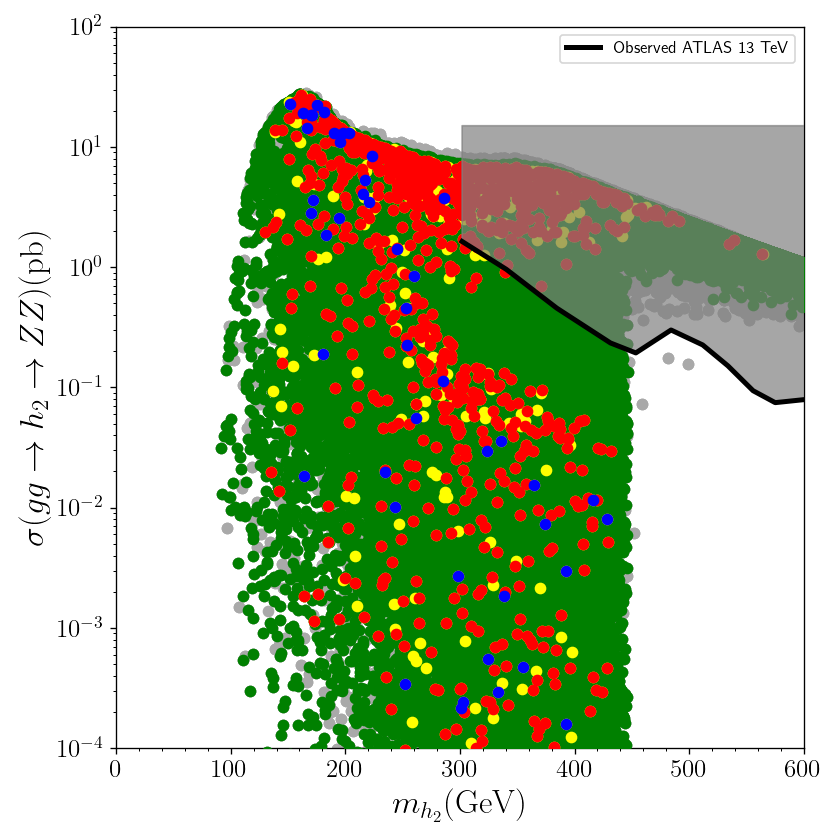}}
\caption{The variation of the cross-sections $\sigma(gg \rightarrow h_2 \rightarrow \tau\tau)$ (top-left), $\sigma(gg \rightarrow h_2 \rightarrow WW)$ (top-right) and $\sigma(gg \rightarrow h_2 \rightarrow ZZ)$ (bottom) with respect to $m_{h_2}$. The color scheme is same as Fig. \ref{fig:Hmasses_1}, and black shaded areas are excluded by Refs. \cite{ATLAS:2020zms,ATLAS:2017jag,ATLAS:2017otj}}
\label{fig:Hmasses_2}
\end{figure}

In this section, the mass spectrum of the LS-2HDM will be analyzed by imposing the aforementioned groups of constraints from theoretical and experimental considerations. First, the correlation between the masses of the LS-2HDM scalars and $\tan\beta$ are illustrated in Fig. \ref{fig:Hmasses_1}. It is seen from Fig. \ref{fig:Hmasses_1} (top-left) that solutions satisfying the theoretical constraints specified by G1 do not introduce further limitations on $m_{h_2}$. The green points in this plot represent solutions that meet both theoretical and experimental constraints specified in groups G1 and G2 within the ranges of $1.6 \lesssim \tan\beta \lesssim 40.0$ and $70 \lesssim m_{h_2} \lesssim 600$ GeV. It's evident that the application of G2 does not result in a significant reduction of the parameter space. The yellow points signify solutions that additionally fulfill the condition $m_{h_1} = 125 \pm 2$ GeV (G1, G2, and G3). While the application of G3 does not restrict the range of $\tan\beta$, it does reduce the number of solutions in the parameter space, and imposes the condition $m_{h_2} \gtrsim 126$ GeV. The solutions indicated by the blue points in Fig.\ref{fig:Hmasses_1} (top-left) depict the parameter space that fulfills all four constraints outlined by groups G1, G2, G3, and G4, i.e. additionally satisfying restrictions coming from $M_W^{\rm CDF}$, which limits the mass of the CP-even scalar boson as $150 \lesssim m_{h_2} \lesssim 430$ GeV. The solutions indicated by the red points depict the parameter space that fulfills G1, G2, G3, and G5. The condition G5 does not restrict the range of parameter space. The region between the vertical dashed lines in Fig. \ref{fig:Hmasses_1} (top-left) corresponds to the case where $h_2$ is the SM-like Higgs boson, which does not satisfy G4, i.e. $W-$boson mass condition. Therefore, the selection of $h_1$ as the SM-like Higgs boson, as specified in constraint G3, is clarified. On the other hand, there is a few solutions satisfying G5 condition. Since it is possible to find solutions compatible with the $W-$boson mass measured by CMS (represented by red points) in almost every region of the parameter space, the G5 condition will not be emphasized in the following sections.

Additionally, the relationship between $m_A$ and $\tan\beta$ is depicted in Fig. \ref{fig:Hmasses_1} (top-right), together with the experimental constraints from ATLAS 13 TeV results \cite{ATLAS:2020zms}, where $m_A$ span the full $80 \lesssim m_A \lesssim 700$ GeV range. It is observed that the application of constraints G1 and G2 imposes limitations as  $80 \lesssim m_A\lesssim  600 + 4.1 \tan\beta$ GeV. Moreover, adding the condition outlined as G3 reduces the number of solutions within the aforementioned regions. However, imposing W-mass constraints described as G4 require $270 \lesssim m_A \lesssim 610$ GeV. The black line indicates the upper exclusion limit from ATLAS 13 TeV analysis \cite{ATLAS:2020zms}, and it eliminates the majority of the solutions in the parameter space. Consequently, the solutions satisfying all constraints outlined as G1, G2, G3 and G4, and also imposing limits from ATLAS 13 TeV model independent scalar mass analysis, the mass of the CP-odd scalar boson and $\tan\beta$ are restricted as $340 \; \text{GeV}\lesssim m_A \lesssim 630 \; \text{GeV}~ \text{and}\ \ \tan\beta\lesssim 8.0$. However, these experimental analyses have been performed for the cases in which the heavy Higgs bosons interact with the quarks and leptons considerably at any value of $\tan\beta$. On the other hand, the models in the LS-2HDM class may not exhibit such a feature, since the interactions between the quarks and these heavy Higgs bosons are suppressed by large $\tan\beta$. In this context, even though the ATLAS analyses result in an exclusion, the solutions accumulated in the faded region in top-left plot of Fig. \ref{fig:Hmasses_1} can still be consistent with the experiments. For the charged scalars of LS-2HDM, the variation of $m_{H^\pm}$ on the $\tan\beta - m_{H^\pm}$ plane is given in Fig. \ref{fig:Hmasses_1} (bottom-left). After applying constraints G1 and G2, the remaining solutions bounded as $110\ \text{GeV} \lesssim m_{H^\pm} \lesssim \ 605 + 3.8 \tan\beta \ \text{GeV}$. However, after applying G3, the lower bound on $m_{H^\pm}$ increases to 150 GeV. After adding cuts from G4, the acceptable solutions satisfy $320 \lesssim m_{H^\pm} \lesssim 630$ GeV, and number of valid solutions are reduced. The exculusion limits from ATLAS 13 TeV \cite{ATLAS:2021upq} requires $\tan\beta\gtrsim 2.0$. 

Until now, the CP-even scalar boson $h_1$ has been identified as the SM-like Higgs boson, and this requirement was described as condition G3. In order to explore the behavior of $m_{h_1}$, we remove condition G3, allowing $m_{h_1}$ to vary without this constraint. The relationship between $\tan\beta$ and $m_{h_1}$, in this case, is presented in Fig. \ref{fig:Hmasses_1} (bottom-right). It is observed that $m_{h_1}$ exhibits distinct behavior in two specific regions of $\tan\beta$. For $\tan\beta\lesssim5$, $m_{h_1}$ shows an inverse proportionality to $\tan\beta$, ranging from $50 \lesssim m_{h_1} \lesssim 288$ GeV, where upper limit arises from condition G4. Conversely, for $\tan\beta\gtrsim 5$, $m_{h_1}$ demonstrates a direct proportionality to $\tan\beta$, with values ranging as $50 \lesssim m_{h_1} \lesssim 450$ GeV. This contrasting behavior in the low and high $\tan\beta$ regions is evident from Eqn. \ref{extraHiggsMasses}.

Moreover, an analysis conducted by ATLAS at 13 TeV has established an upper limit on $m_{h_2}$ through various channels, including the decay modes of $h_2$ into two tau-leptons, two $W-$bosons, or two $Z-$bosons \cite{ATLAS:2020zms,ATLAS:2017jag,ATLAS:2017otj}. To investigate these limitations, the variation of the cross-sections $\sigma(gg \rightarrow h_2 \rightarrow \tau\tau)$, $\sigma(gg \rightarrow h_2 \rightarrow WW)$ and $\sigma(gg \rightarrow h_2 \rightarrow ZZ)$ with respect to $m_{h_2}$ are depicted in Fig. \ref{fig:Hmasses_2}. It is observed from Fig. \ref{fig:Hmasses_2} (top-left) that as $m_{h_2}$ increases, the production cross-section in channel $\sigma(gg \rightarrow h_2 \rightarrow \tau\tau)$ decreases exponentially. According to the restrictions outlined in Ref. \cite{ATLAS:2020zms}, nearly half of the solutions that satisfy criteria G4 are excluded, and the allowed parameter range necessitates $m_{h_2} \gtrsim 200$ GeV, along with $\sigma(gg \rightarrow h_2 \rightarrow \tau\tau) \lesssim 2\times 10^{-1}$ pb. Observing in Fig. \ref{fig:Hmasses_2} (top-right), it's easily seen that the $\sigma(gg \rightarrow h_2 \rightarrow WW)$ value decreases as $m_{h_2}$ increases, as expected. Importantly, the solutions that satisfy G4 remain unaffected by the constraint reported in Ref. \cite{ATLAS:2017jag}. For $m_{h_2}\gtrsim 450$ GeV, solutions with $\sigma(gg \rightarrow h_2) \simeq 1.0$ pb and $BR(h_2 \rightarrow WW) \simeq 1.0$ can be obtained.  A similar behavior can be seen in Fig. \ref{fig:Hmasses_2} (bottom) where the dependence of $\sigma(gg \rightarrow h_2 \rightarrow WW)$ on $m_{h_2}$ is plotted, the production cross-section decreases as $m_{h_2}$ increases. No solution satisfying the G4 condition is excluded by the constraints outlined in Ref. \cite{ATLAS:2017otj}. It should be noted that for $m_{h_2}$ below 300 GeV, the cross-section for the decay channel $\sigma(gg \rightarrow h_2 \rightarrow ZZ)$ remains above 1 pb. After considering these findings from Fig.\ref{fig:Hmasses_2}, the mass of the heavier CP-even scalar boson is restricted as $150 \lesssim m_{h_2} \lesssim 400$ GeV.

\subsection{Lepton Flavor Universality in Z boson Decay}

\begin{figure}[t!]
\includegraphics[scale=0.4]{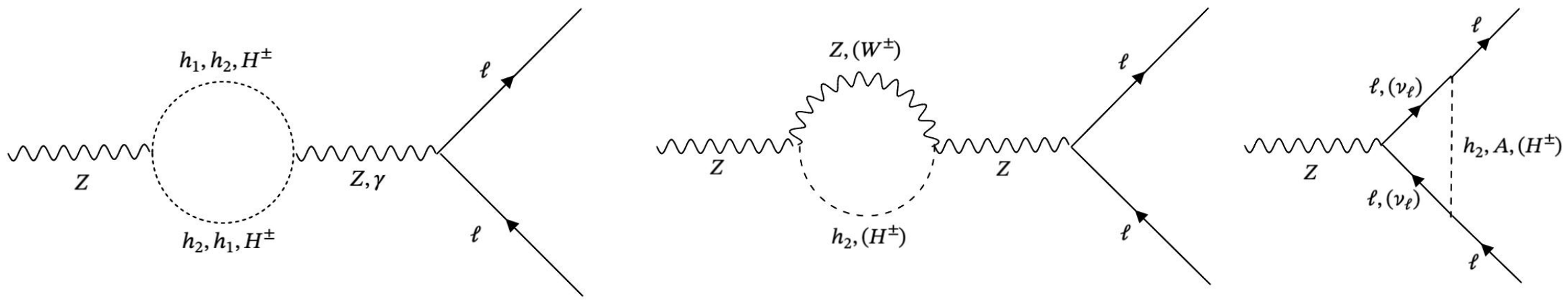}
\caption{The diagrams illustrating the contributions to LFU from the new Higgs bosons through decays of $Z-$boson.}
\label{fig:Zdiags}
\end{figure}

Another set of constraints on the solutions arises from the decay proceeses involving $Z-$boson and $\tau-$lepton. Even though such processes can yield LFU in SM itself due to the mass hierarchy among the fermion families, the experimental measurements reveal results beyond the SM predictions. In this context, the new contributions to LFU can be utilized through the processes involving the extra Higgs bosons. Their contributions are displayed diagrammatically in Fig.  \ref{fig:Zdiags}. The couplings of $Z-$boson with the fermions and the extra Higgs bosons can be given as \cite{Chun:2016hzs}

\begin{equation}
\begin{aligned}
-{\cal L} &= {g\over \cos\theta_W} Z^\mu \big\{ \bar{f} \gamma_\mu (((T_3(f_{L}) -Q(f_{L}) \sin\theta_W^2)\delta g_{L}) P_L  + ((T_3(f_{R}) Q(f_{R}) \sin\theta_W^2) +\delta g_{R}) P_R) f \\
&+i (-{1\over2} + \sin\theta_W^2) H^+ \overleftrightarrow{\partial_\mu}  H^- + A \overleftrightarrow{\partial_\mu} H \big\}.
\end{aligned}
\label{eq:lagZleptonH}
\end{equation}
where the contributions from the new physics can be seen through $\delta g_{R,L}$, which can be calculated as follows

\begin{equation}
\begin{aligned}
\delta g_L^{\rm 2HDM} &= k\left[-\dfrac{1}{2}B_Z\left(r_A\right)-\dfrac{1}{2}B_Z\left(r_{h_2}\right)-2C_Z\left(r_A,r_{h_2}\right) +  \sin^2\theta_W \left( B_Z\left(r_A\right) + B_Z\left(r_{h_2}\right) + \tilde{C}_Z(r_A) + \tilde{C}_Z(r_{h_2})\right) \right], \\
\delta g_R^{\rm 2HDM} &= k\left[2C_Z\left( r_A, r_{h_2}\right)  -2C_Z\left( r_{H^\pm},r_{H^\pm}\right) +\tilde{C}_Z(r_{H^\pm})  - \dfrac{1}{2}\tilde{C}_Z(r_A)-\dfrac{1}{2}\tilde{C}_Z(r_{h_2}) \right.,  \\
 &+ \left.  \sin^2\theta_W\left( B_Z\left(r_A\right) +  B_Z\left(r_{h_2}\right) + 2B_Z\left(r_{H^\pm}\right) +\tilde{C}_Z(r_A) + C_Z\left( r_{H^\pm},r_{H^\pm}\right) \right) \right ], 
\end{aligned}
\label{eq:glgr}
\end{equation}
with $k=m_f^2 \tan^2\beta / 16\pi^2 v^2$,  $m_f$ are fermion masses, $\Lambda$ is the cut-off scale, $\mu$ is the renormalization scale, and  $r_\phi = m_\phi^2 / m_Z^2$ ($\phi = A, h_2, H^\pm$). The loop functions employed in Eqn. \ref{eq:glgr} are calculated as

\begin{equation}
\begin{aligned}
B_Z(r)&= - \dfrac{1}{2}\left(\ln\Lambda^2 +\log(4\pi) \right)-\dfrac{1}{4} + \dfrac{1}{2}\log r  ,\\
\tilde{C}_Z(r)&= - \dfrac{1}{2}\left(\ln\Lambda^2 +\log(4\pi) \right)-\dfrac{1}{2}  -r \left(1+\log r \right) +r^2\left(\log r \log(1+r^{-1}) - {\rm dilog}(-r^{-1}) \right) \\
&- \dfrac{i\pi}{2}\left(1-2r+2r^2\log\left(1+r^{-1}\right) \right) , \\
C_Z\left(r_1,r_2\right) &= C_Z(\dfrac{m_{\phi_1}^2}{m_{Z}^2},\dfrac{m_{\phi_2}^2}{m_{Z}^2}) = C_{00}(0,0,m_Z^2,m_{\phi_1}^2,m_{\phi_2}^2), \\
C_{00} &= \dfrac{1}{4}\left(\ln\Lambda^2 + \log(4\pi)-\ln(\dfrac{m_Z^2}{\mu^2})\right)  + \dfrac{3}{8}-\dfrac{r_1^2\ln r_1}{4(r_1-1)(r_1-r_2)}
+ \dfrac{r_2^2\ln r_2}{4(r_2-1)(r_1-r_2)}.
\end{aligned}
\label{eq:loopfunc}
\end{equation}

The deviation from LFU in the models can be parametrised with $\delta g_{R,L}$ as follows

\begin{equation}
\begin{aligned}
 \delta_{ll} = {2 g_L^e{\rm Re}(\delta g^{\rm LS-2HDM}_L)+ 2 g_R^e{\rm Re}(\delta g^{\rm LS-2HDM}_R) \over {g_L^e}^2 + {g_R^e}^2 },
\end{aligned}
 \label{eq:deltall}
\end{equation}
where $l=\mu,\tau$.
Throughout this analysis, lepton universality in the $Z-$boson decay are investigated through partial decay width of $Z-$boson into the leptons, which can be summarized with the relevant experimental measurements \cite{Angel:2013hla,Chun:2016hzs} as follows:

\begin{equation}
\begin{aligned}
{\Gamma_{Z\to \mu^+ \mu^-}\over \Gamma_{Z\to e^+ e^- }} &=& 1.0009 \pm 0.0028
\,,  \\
{\Gamma_{Z\to \tau^+ \tau^- }\over \Gamma_{Z\to e^+ e^- }} &=& 1.0019 \pm 0.0032
\,.
\end{aligned} \hspace{0.3cm} \Rightarrow \hspace{0.3cm} \left\lbrace \begin{aligned}
\delta_{ \mu \mu} &=& {\Gamma_{Z\to \mu^+ \mu^-}\over \Gamma_{Z\to e^+ e^- }}  - 1 = (9 \pm 28 )\times10^{-4}, \\
\delta_{ \tau\tau} &=& {\Gamma_{Z\to \tau^+ \tau^-}\over \Gamma_{Z\to e^+ e^- }}  - 1 = (19 \pm 32 )\times10^{-4},
\end{aligned}\right.
 \label{lu-zdecay}
\end{equation}
where $\delta_{\mu\mu}$ and $\delta_{\tau\tau}$ are the parameters defined to measure the deviation from LFU.

\begin{figure}[t!]
\centering
\subfigure{\includegraphics[scale=0.35]{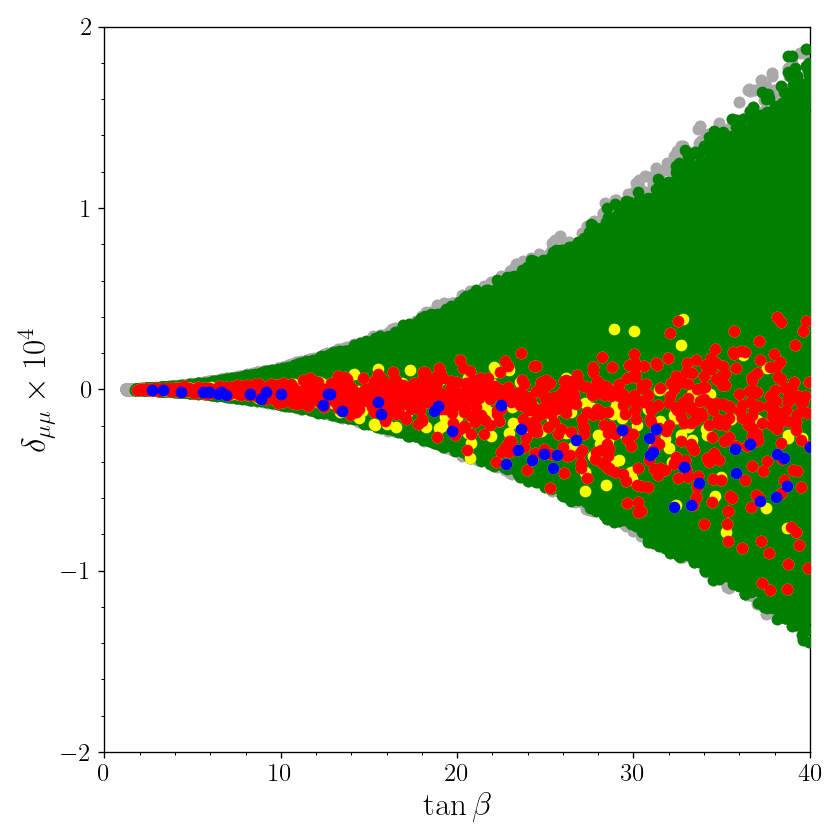}}%
\subfigure{\includegraphics[scale=0.35]{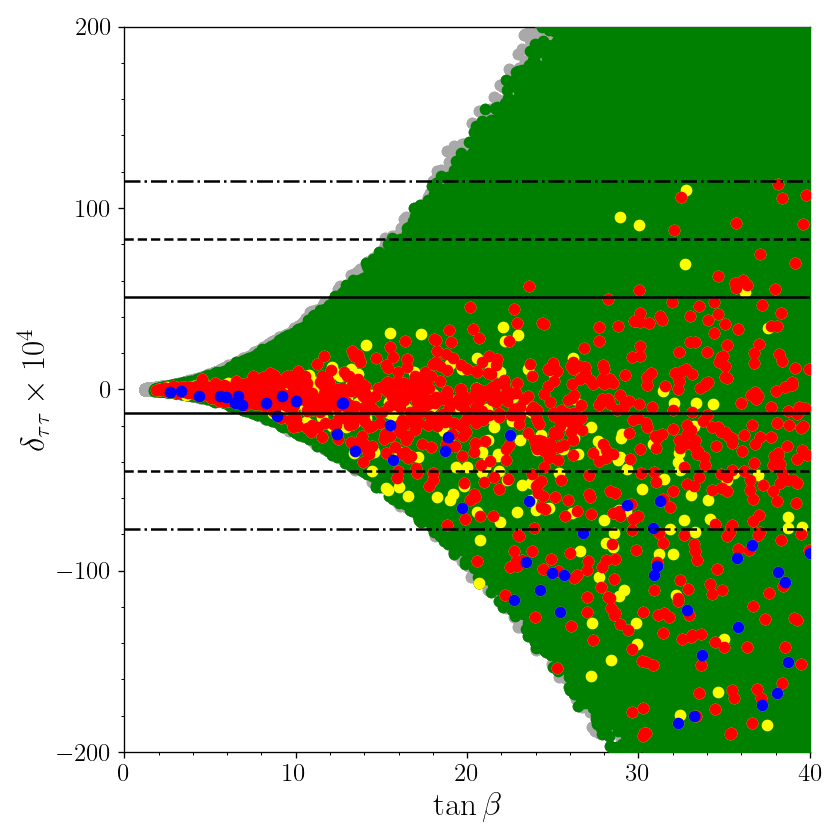}}
\caption{The variation of $\delta_{\mu\mu}$ with respect to $\tan\beta$ (left) and $\delta_{\tau\tau}$ with respect to $\tan\beta$ (right). Color coding of the plots is the same as Fig. \ref{fig:Hmasses_1}.  The solid, dashed and dash dotted lines indicate deviations from the current limits by Eqn. \ref{lu-zdecay} within $1\sigma$, $2\sigma$, and $3\sigma$, respectively. For $\delta_{\mu\mu}$ (left), all solutions lie within 1$\sigma$ vicinity.}
\label{fig:LFU_Z}
\end{figure}

In order to understand the impact of LS-2HDM on lepton universality, such as the loop-induced contributions to the $Z-$boson decay widths, it's essential to explore the precise measurements of lepton universality ratios provided in Eqn. \ref{lu-zdecay} within the framework of LS-2HDM. To this end, the distribution of solutions that satisfy the constraints outlined as G1, G2, G3 and G4 are plotted on  $\delta_{ll}$ and $\tan\beta$ plane in left for $\delta_{\mu\mu}$ and in right for $\delta_{\tau\tau}$, in Fig.  \ref{fig:LFU_Z}.

It is seen from Fig. \ref{fig:LFU_Z} that the dependence of $\delta_{ll}$ on $\tan\beta$ is quadratic as indicated in Eqn. \ref{eq:glgr}, in both positive or negative directions depending on the scalar masses of the LS-2HDM. Moreover, Fig. \ref{fig:LFU_Z} (left) demonstrates that the deviation of $\delta_{\mu\mu}$ from the LFU lies within $1\sigma$ level and increases with the $\tan\beta$. Notably, to satisfy the $W-$boson mass constraint given by G4, the value of $\delta_{\mu\mu}$ must be negative. Furthermore, Fig. \ref{fig:LFU_Z} (right) demonstrates that $\delta_{\tau\tau}$ displays a greater sensitivity to $\tan\beta$, due to the contribution of $m_{\tau}^2$. It is important to note that the contribution can be either positive or negative, but if the parameter space is restricted by G4, $\delta_{\tau\tau}$ is typically negative with a few exceptions. The majority of solutions satisfying all constraints—G1, G2, G3, and G4—are observed to fall within a $3\sigma$ deviation from the observed value of $\delta_{\tau\tau}=1.0019$.

\begin{table}[ht!]
    \centering
\renewcommand{\arraystretch}{1.5} 
\setlength{\tabcolsep}{16pt}
    \begin{tabular}{|c|c|c|c|}
\hline
$\delta_{\tau\tau}$ & $-12.9518 \times 10^{-4} \; (\leq 1\sigma)$ & $ -34.1836 \times 10^{-4} \; (\leq 2\sigma) $ & $-63.6799 \times 10^{-4} \; (\leq 3\sigma)$ \\
\hline
$\tan\beta$ & $12.774$ & $13.45$ & $29.332$ \\
\hline
$m_{h_1}$ & $126.901$ & $125.808$ & $126.263$ \\
\hline
$m_{h_2}$ & $261.613$ & $234.808$ & $364.553$ \\
\hline
$m_{A}$ & $350.786$ & $558.726$ & $451.389$ \\
\hline
$m_{H^\pm}$ & $393.202$ & $ 578.161$ & $500.79$ \\
\hline
$m_{3}^2$ & $-4327.4$ & $-4004.4$ & $-4518.6$ \\
\hline
$|\cos(\beta - \alpha)|$ & $0.0497$ & $0.0455$ & $0.0591$\\
\hline
$M_{\rm W}^{\rm LS-2HDM} - M_{\rm W}^{\rm CDF} $ & $-0.0309 \; (3.28 \sigma)$ & $-0.0241 \; (2.56 \sigma)$ & $-0.0227  \; (2.41 \sigma) $  \\
\hline
$M_{\rm W}^{\rm LS-2HDM} - M_{\rm W}^{\rm CMS}$ & $0.0424 \; (4.61 \sigma)$ & $ 0.0492 \; (5.35 \sigma)$ & $0.0506 \; (5.50 \sigma)$ \\
\hline
    \end{tabular}
    \caption{Properties of solutions which are selected from Fig. \ref{fig:LFU_Z} (right). All masses are given in GeV unit. $1\sigma,2\sigma,3\sigma$ are the deviations from value of $\delta_{\tau\tau}$ (1.0019). All the points are selected to yield possible minimum values for $\cos(\beta - \alpha)$.}
    \label{tab:LFU_Z}
\end{table}

To exemplify the behaviour of $\delta_{\tau\tau}$ we have listed three benchmark points in Table \ref{tab:LFU_Z}. These points are selected to be consistent all the constraints applied so far, and they yield possible minimum value for $\cos(\beta - \alpha)$ for each vicinity of the experimental measurements on LFU. The last two rows in the table show the accommodation of $W-$boson mass in terms of the differences and uncertainty with respect to CDF and CMS results, respectively. These solutions reveal that the heavy Higgs bosons should weigh in the mass range from 250 GeV to about 500 GeV. These solutions together with the blue points in the left panel of Fig. \ref{fig:LFU_Z}reveal that the CDF compatible solutions yield also large deviations in LFU processes, and one can accommodate CDF result only up to about $2\sigma$ consistently.

\subsection{Lepton Flavor Universality in Tau Decay}

\begin{figure}[h!]
\includegraphics[scale=0.4]{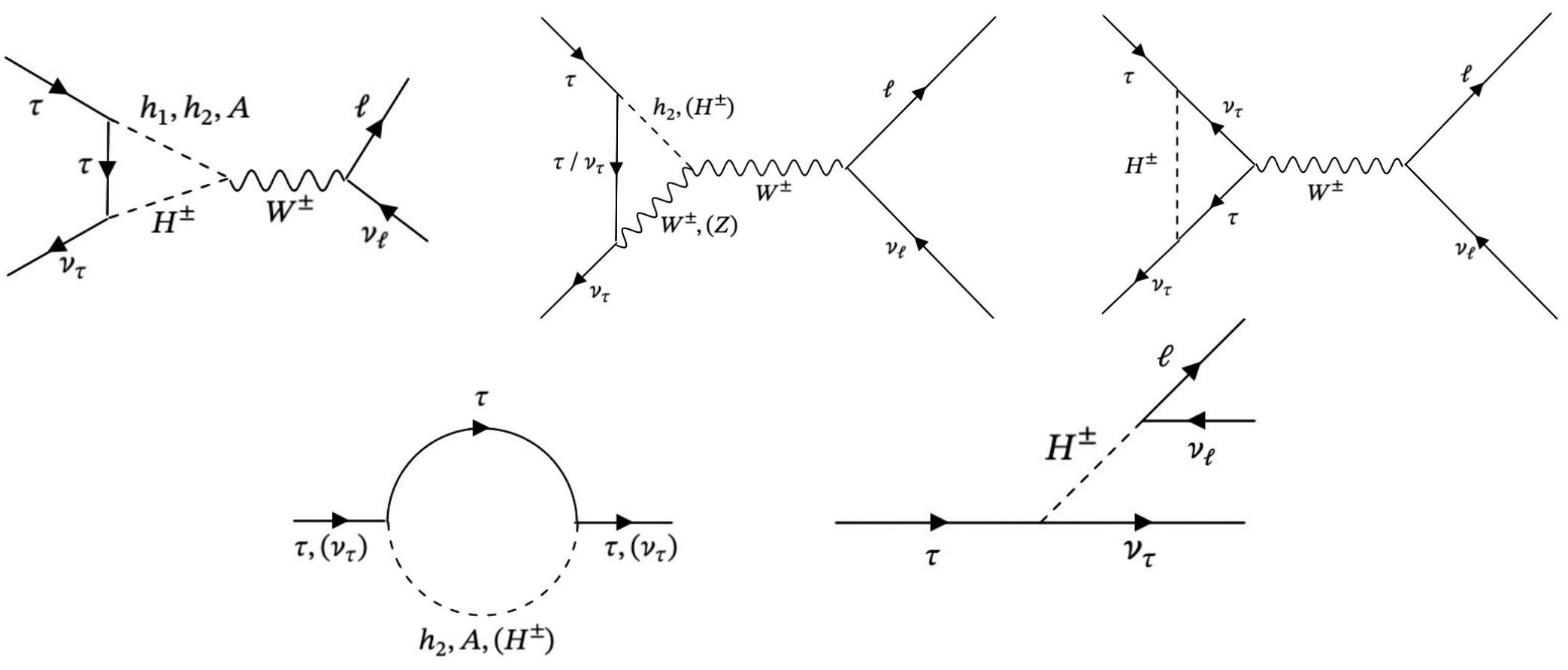}
\caption{The diagrams illustrating the contributions to LFU from the new Higgs bosons through decays of $\tau$.}
\label{fig:taudiags}
\end{figure}

The deviations from LFU can also be tested in decays of $\tau-$lepton. The recent analyses \cite{HFLAV:2022esi} consider the pure leptonic and semi-hadronic decay processes. Since the QCD sector remains intact in LS-2HDM, and the new Higgs bosons can interfere these processes through the electroweak and Yukawa interactions, we consider only the pure leptonic decays to calculate the deviations from LFU. The relevant diagrams are displayed in Fig. \ref{fig:taudiags} The deviation in $\tau$ decays can be analysed by considering the ratio of the gauge couplings of different lepton families as follows:

\begin{equation}
\left( g_\tau \over g_\mu \right) = 1+ \delta_{\rm loop}, \qquad  \left( g_\tau \over g_e \right) =1+ \delta_{\rm tree}+ \delta_{\rm loop}, \qquad \left( g_\mu \over g_e \right) = 1+ \delta_{\rm tree},
\label{deltas-data}
\end{equation}
where $\delta_{\rm loop}$ and $\delta_{\rm tree}$ are the contributions to SM value from loop and tree level processes. In LS-2HDM, these contributions are calculated as \cite{Abe:2015oca,Chun:2016hzs,Wang:2018hnw}

\begin{equation}
\begin{aligned}
\delta_{\rm tree} &={m_\tau^2 m_\mu^2 \over 8 m^4_{H^\pm}} t^4_\beta
- {m_\mu^2 \over m^2_{H^\pm}} t^2_\beta {g(m_\mu^2/m^2_\tau) \over f(m_\mu^2/m_\tau^2)} , \\
\delta_{\rm loop} &= {1 \over 16 \pi^2} { m_\tau^2 \over v^2}  t^2_\beta \left[1 + {1\over4} \left( H(m_A^2/m_{H^\pm}^2) + s^2_{\beta-\alpha} H(m_{h_2}^2/m_{H^\pm}^2)  + c^2_{\beta-\alpha} H(m_{h_1}^2/m_{H^\pm}^2\right) \right],
\end{aligned}
\label{deltas}
\end{equation}
where

\begin{align}
  f(x)&\equiv 1-8x+8x^3-x^4-12x^2 \ln(x), \nonumber \\ 
  g(x)&\equiv 1+9x-9x^2-x^3 +6x(1+x)\ln(x),\\
  H(x) &\equiv \ln(x) (1+x)/(1-x), \nonumber  
\end{align}
and $t_{\beta} = \tan\beta $, $s_{\beta-\alpha} = \sin(\beta -\alpha)$, $c_{\beta-\alpha} = \cos(\beta -\alpha)$ are defined for brevity. Using pure leptonic processes, HFLAV collaboration obtained the values of the ratios given in Eqn. \ref{deltas} as \cite{HFLAV:2022esi} 

 \begin{equation}
\left( g_\tau \over g_\mu \right) =  1.0009 \pm 0.0014, \qquad  \left( g_\tau \over g_e \right) = 1.0027 \pm 0.0014, \qquad \left( g_\mu \over g_e \right)  = 1.0019 \pm 0.0014 .
\label{hfag-data}
\end{equation}
 These averages define the ranges for loop and tree-level contributions as
 
 \begin{align}
\delta_{loop}=0.0009\pm0.0014~\ \ \text{and}\ \ \ \ \delta_{tree}=0.0019\pm0.0014~,
 \label{hfag-deltas}\end{align}
and these limits will be considered for further analysis.

\begin{figure}[t!]
\centering
\subfigure{\includegraphics[scale=0.35]{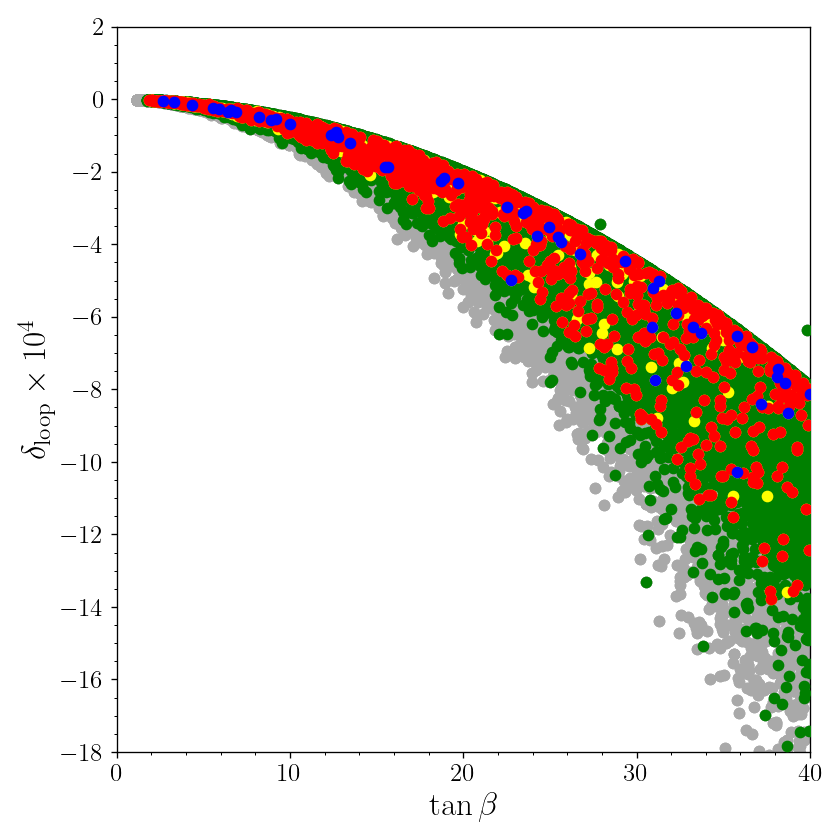}}
\subfigure{\includegraphics[scale=0.35]{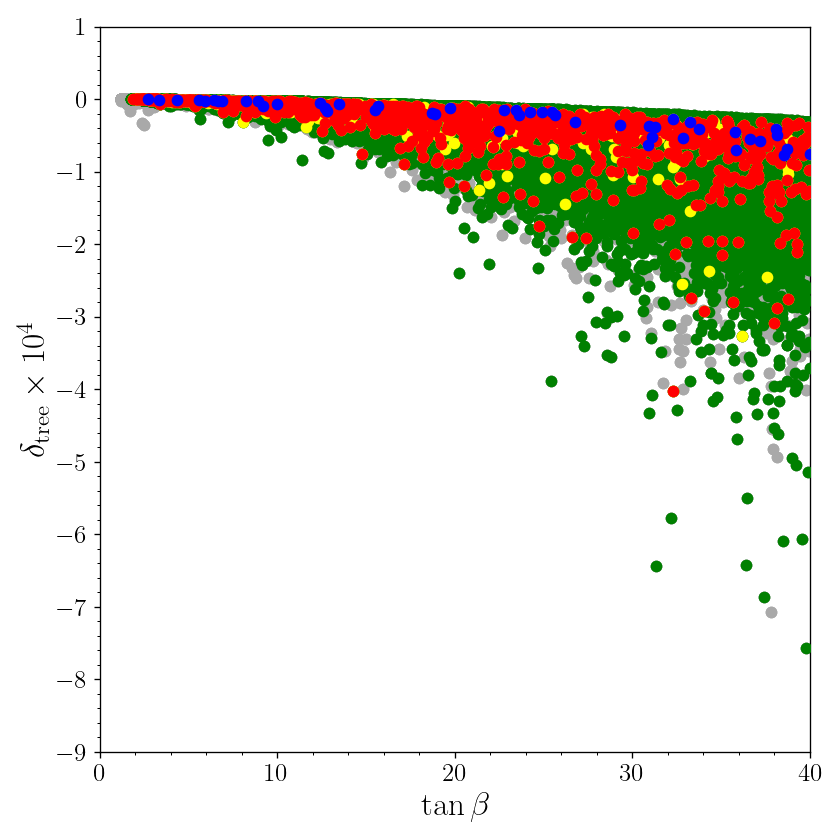}}
\caption{$ \delta_{loop}$ vs $\tan\beta$ (left), and $\delta_{tree}$ vs $\tan\beta$ (right). Color coding is the same as Fig. \ref{fig:Hmasses_1}. All solutions lay within the $1\sigma$ vicinity of HFLAV results.}
\label{fig:LFU_tau}
\end{figure}

To examine the relationship between solutions satisfying all four constraints, they are plotted in the $\delta_{loop}$ and $\delta_{tree}$ versus $\tan\beta$ planes in Fig. \ref{fig:LFU_tau} (left and right), respectively. It is observed from Fig. \ref{fig:LFU_tau} that applying constraints reduces the number of solutions, and restricts them in the negative plane. Furthermore, all solutions shown in red in Fig. \ref{fig:LFU_tau} (left) lie within $2\sigma$ vicinity of the HFVAL average value of $\delta_{loop}$ given in Eqn. \ref{hfag-deltas}. Remarkably, for solutions with $\tan\beta\lesssim20$, this proximity is confined within the 1$\sigma$ range. On the other hand, concerning $\delta_{tree}$, solutions meeting all constraints are clustered just below zero, confined within a $1\sigma$ to $2\sigma$ window of the average value, as observed in Fig. \ref{fig:LFU_tau} (right). Finally, it is also apparent from Fig. \ref{fig:LFU_tau} that with increasing $\tan\beta$, loop contributions to the ratios from LS-2HDM become more significant in comparison to those at the tree level.pl

\subsection{W boson Mass in LS-2HDM}
\label{SectionD}

\begin{figure}[h!]
\includegraphics[scale=0.4]{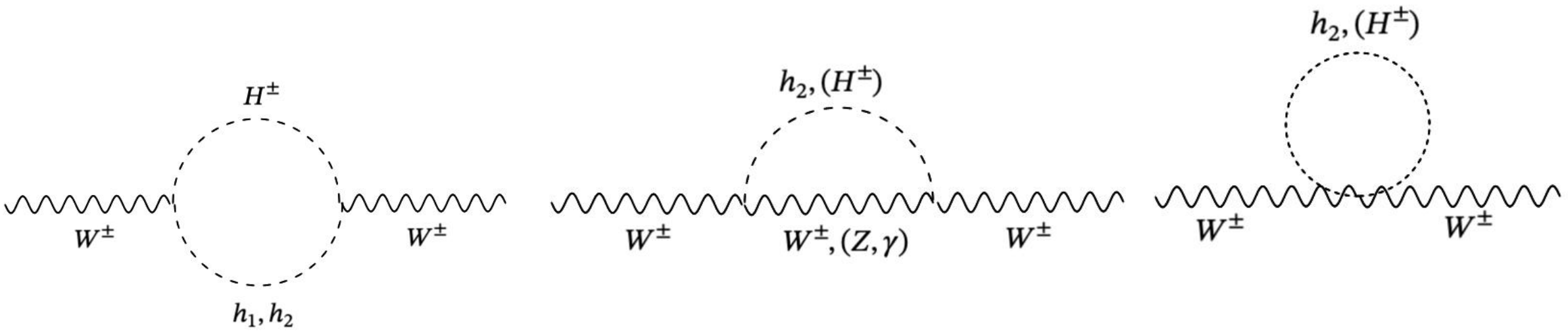}
\caption{The diagrams illustrating the contributions to $W-$boson mass from the new Higgs bosons.}
\label{fig:Wdiags}
\end{figure}

As being the main motivation of our study, the new scalar states alter the $W-$boson mass through loops, since they are non-trivially charged under the SM gauge group in 2HDMs. Their contributions can be analyses through the self-energies of the gauge bosons. These scalar states can arise these contributions through their interactions with $W-$boson, and the mixing among them. These contributions are illustrated in Fig. \ref{fig:Wdiags}. Even though one can constrain these contributions by fitting the $W-$boson mass with CDF and/or CMS measurements, also the results from the analyses over the oblique parameters should be applied and satisfied. These oblique parameters can be defined by considering the self-energies of $W-$boson, $Z-$boson and also contributions to mixing of these gauge bosons yielding $Z\gamma$ processes \cite{Peskin:1990zt,Peskin:1991sw,Haber:2010bw}. In addition, the contributions to $W-$boson mass can also be written in terms of these oblique parameters as follows \cite{Heinemeyer:2022ith}

\begin{align}
M_W^2 = {M_W^{\rm SM}}^2
\left(
1 + \frac{s_W^2}{c_W^2 - s_W^2} \Delta r'
\right) \ ,
\label{MW-STU}
\end{align}
with the loop contributions to the two-point function
\begin{align}
\Delta r' = \frac{\alpha}{s_W^2}
\left(
-\frac{1}{2} S +
c_W^2 T +
\frac{c_W^2 - s_W^2}{4 s_W^2} U
\right) \ ,
\label{Deltar-STU}
\end{align}
which represents the measure of deviations from SM, and where $s_W = \sin\theta_W$ and $c_W = \cos\theta_W$ with $\theta_W$ is being the weak angle. In Eqn. \ref{Deltar-STU}, $S$, $T$, and $U$ denote the oblique parameters, containing the effects of incorporation of additional scalar bosons contributions through loops within the LS-2HDM framework to $M_W$. These parameters are defined based on precision measurements in electroweak physics, and in SM, $S=T=U=0$ serves as the reference point.  The explicit form of the $S$, $T$, $U$ parameters are given as \cite{Haber:2010bw}

\begin{equation}
\begin{aligned}
S=&\frac{1}{\pi m_Z^2}\left\{-\mathcal{B}_{22}\left(m_Z^2 ; m_{H^{ \pm}}^2, m_{H^{ \pm}}^2\right)+\sin ^2(\beta-\alpha) \mathcal{B}_{22}\left(m_Z^2 ; m_{h_2}^2, m_{A}^2\right)\right. \\
 &+\cos ^2(\beta-\alpha)\left[\mathcal{B}_{22}\left(m_Z^2 ; m_{h_1}^2, m_{A}^2\right)+\mathcal{B}_{22}\left(m_Z^2 ; m_Z^2, m_{h_2}^2\right)-\mathcal{B}_{22}\left(m_Z^2 ; m_Z^2, m_{h_1}^2\right)\right. \\
& \left.\left.-m_Z^2 \mathcal{B}_0\left(m_Z^2 ; m_Z^2, m_{h_2}^2\right)+m_Z^2 \mathcal{B}_0\left(m_Z^2 ; m_Z^2, m_{h_1}^2\right)\right]\right\} ,\\
\end{aligned}
\label{S_equation}
\end{equation}

\begin{equation}
\begin{aligned}
 T=&\frac{1}{16 \pi s_W^2 m_W^2}\left\{\mathcal{F}\left(m_{H^{\pm}}^2, m_{A}^2\right)+\sin ^2(\beta-\alpha)\left[\mathcal{F}\left(m_{H^{\pm}}^2, m_{h_2}^2\right)-\mathcal{F}\left(m_{A}^2, m_{h_2}^2\right)\right]\right. \\
& +\cos ^2(\beta-\alpha)\left[\mathcal{F}\left(m_{H^{\pm}}^2, m_{h_1}^2\right)-\mathcal{F}\left(m_{A}^2, m_{h_1}^2\right)+\mathcal{F}\left(m_W^2, m_{h_2}^2\right)-\mathcal{F}\left(m_W^2, m_{h_1}^2\right)\right. \\
& -\mathcal{F}\left(m_Z^2, m_{h_2}^2\right)+\mathcal{F}\left(m_Z^2, m_{h_1}^2\right)+4 m_Z^2 B_0\left(0 ; m_Z^2, m_{h_2}^2\right)-4 m_Z^2 B_0\left(0 ; m_Z^2, m_{h_1}^2\right) \\
& \left.\left.-4 m_W^2 B_0\left(0 ; m_W^2, m_{h_2}^2\right)+4 m_W^2 B_0\left(0 ; m_W^2, m_{h_1}^2\right)\right]\right\}, \\
\end{aligned}
\label{T_equation}
\end{equation}

\begin{equation}
\begin{aligned}
 S+U=&\frac{1}{\pi m_W^2}\left\{\mathcal{B}_{22}\left(m_W^2 ; m_{H^{\pm}}^2, m_{A}^2\right)-2 \mathcal{B}_{22}\left(m_W^2 ; m_{H^{\pm}}^2, m_{H^{\pm}}^2\right)+\sin ^2(\beta-\alpha) \mathcal{B}_{22}\left(m_W^2 ; m_{H^{\pm}}^2, m_{h_2}^2\right)\right. \\
& +\cos ^2(\beta-\alpha)\left[\mathcal{B}_{22}\left(m_W^2 ; m_{h_1}^2, m_{H^{\pm}}^2\right)+\mathcal{B}_{22}\left(m_W^2 ; m_W^2, m_{h_2}^2\right)-\mathcal{B}_{22}\left(m_W^2 ; m_W^2, m_{h_1}^2\right)\right. \\
& \left.\left.+m_W^2 \mathcal{B}_0\left(m_W^2 ; m_W^2, m_{h_2}^2\right)-m_W^2 \mathcal{B}_0\left(m_W^2 ; m_W^2, m_{h_1}^2\right)\right]\right\}, \\
\end{aligned}
\label{SU_equation}
\end{equation}
where
\begin{equation}
\begin{aligned}
&\mathcal{B}_{22}\left(q^2 ; m_{1}^2, m_{2}^2\right) \equiv \mathcal{B}_{22}\left(q^2 ; m_{1}^2, m_{2}^2\right) - \mathcal{B}_{22}\left(0 ; m_{1}^2, m_{2}^2\right), \\
&\mathcal{B}_{0}\left(q^2 ; m_{1}^2, m_{2}^2\right) \equiv \mathcal{B}_{0}\left(q^2 ; m_{1}^2, m_{2}^2\right) - \mathcal{B}_{0}\left(0 ; m_{1}^2, m_{2}^2\right) ,
\end{aligned}
\label{two_loop_func}
\end{equation}
are the loop functions arising from two-point loop integrals given as

\begin{equation}
\begin{aligned}
&B_{22}\left(q^2 ; m_1^2, m_2^2\right)  =\frac{1}{4}(\Delta+1)\left[m_1^2+m_2^2-\frac{1}{3} q^2\right]-\frac{1}{2} \int_0^1 d x X \ln (X-i \epsilon), \\
&B_0\left(q^2 ; m_1^2, m_2^2\right)  =\Delta-\int_0^1 d x \ln (X-i \epsilon),
\end{aligned}
\end{equation}
where
\begin{equation}
X \equiv m_1^2 x+m_2^2(1-x)-q^2 x(1-x), \quad \Delta \equiv \frac{2}{4-d}+\ln 4 \pi-\gamma .
\end{equation}
Shorthand notation $\mathcal{F}$ in Eqn. \ref{T_equation} is defined as
\begin{equation}
\begin{aligned}
\mathcal{F}\left( m_1^2 , m_2^2\right) \equiv \dfrac{1}{2}\left(m_1^2 + m_2^2\right) - \dfrac{m_1^2 m_2^2}{m_1^2 - m_2^2}\ln\left(\dfrac{m_1^2}{m_2^2}\right),
\end{aligned}
\end{equation}
satisfying the following relations
\begin{equation}
\mathcal{F}\left(m_1^2,m_2^2\right) = \mathcal{F}\left(m_2^2,m_1^2\right) \; {\rm and} \; \mathcal{F}\left(m^2,m^2\right) =0.    
\end{equation}

\begin{figure}[t!]
\centering
\subfigure{\includegraphics[scale=0.3]{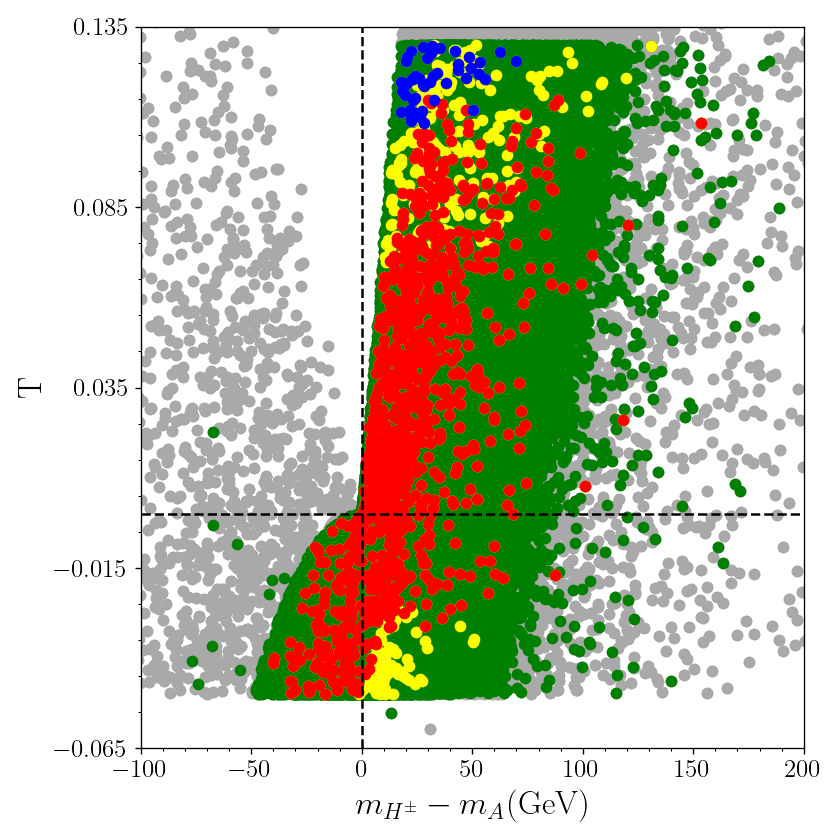}}%
\subfigure{\includegraphics[scale=0.3]{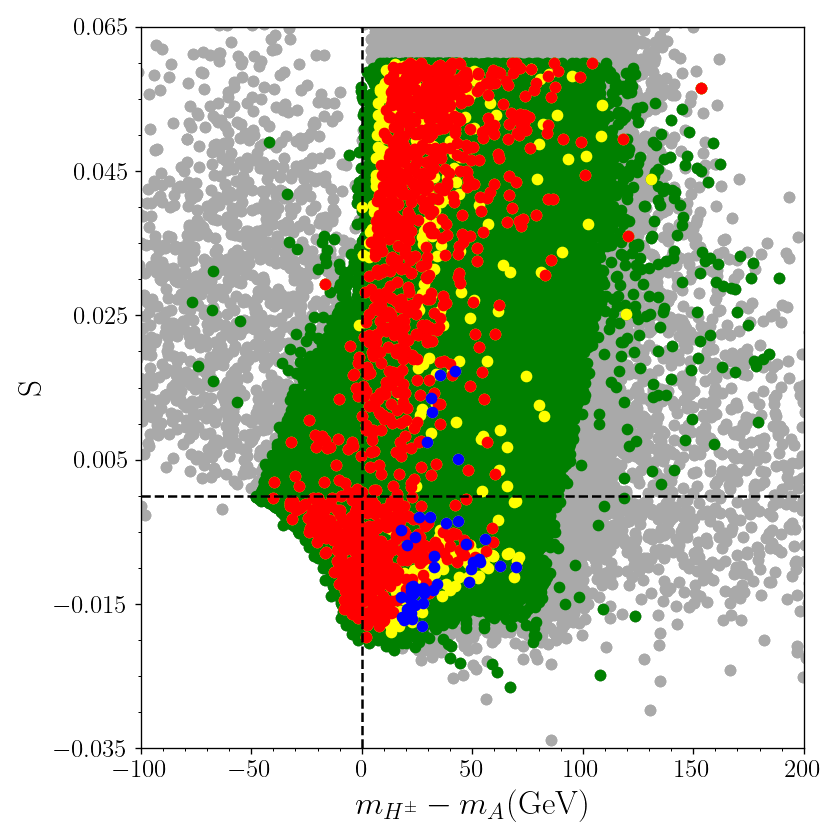}}
\subfigure{\includegraphics[scale=0.3]{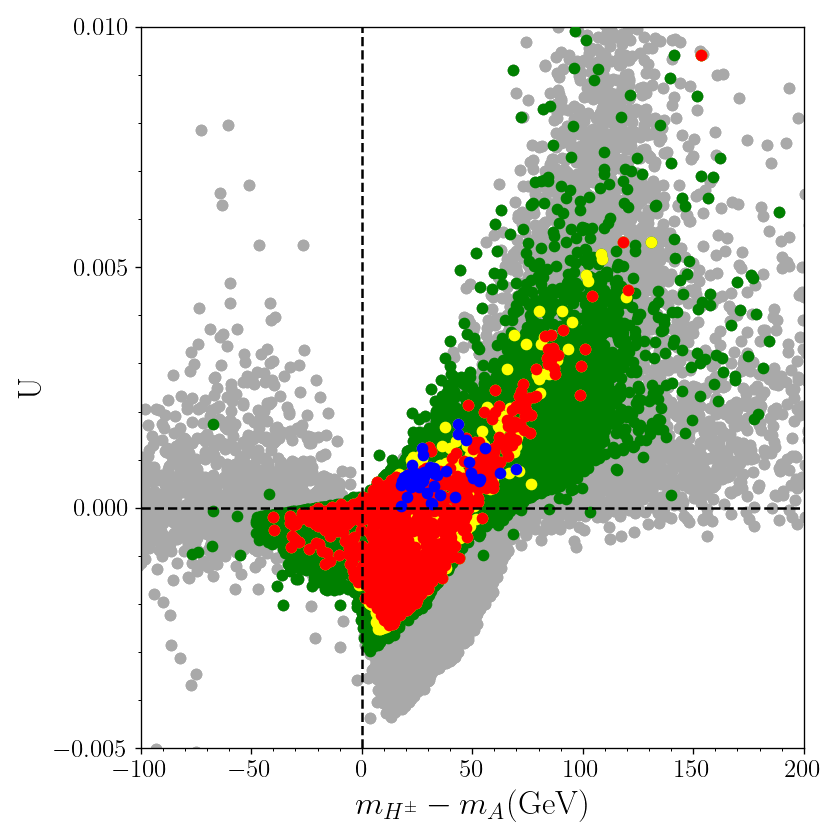}}%
\subfigure{\includegraphics[scale=0.3]{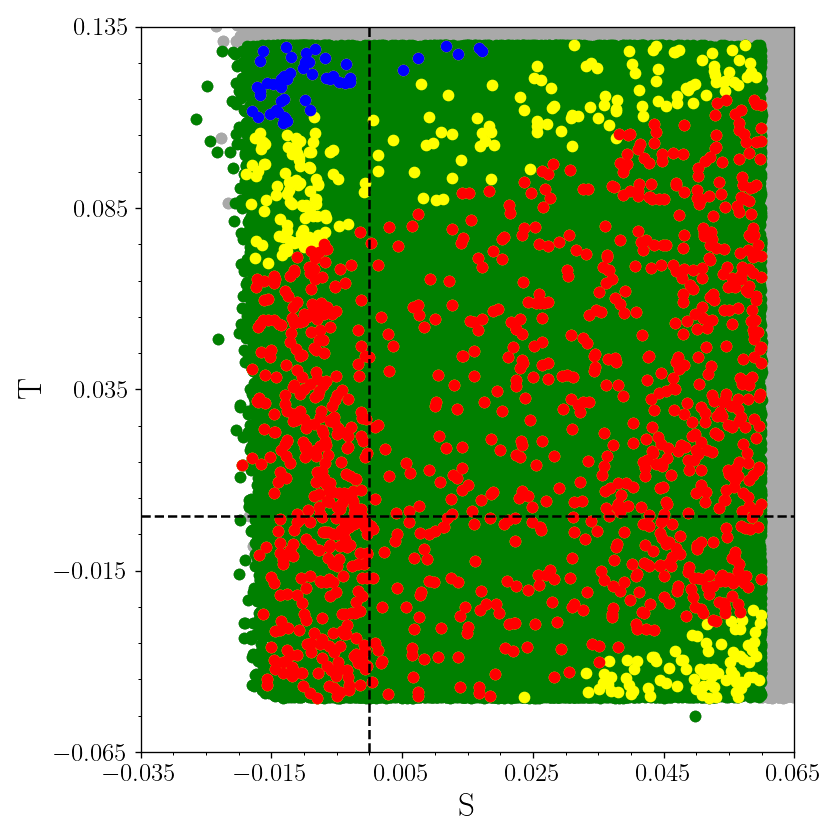}}
\caption{The Oblique parameters in correlation with the mass differences among thr Higgs boson. The bottom-right panel also displays possible correlation between $S$ and $T$ parameters. The color coding is the same as in Fig. \ref{fig:Hmasses_1}. The vertical and horizontal dashed lines indicate the solutions with $S,T,U = 0$ as referred to SM.}
\label{fig:obliques}
\end{figure}

\begin{figure}[t!]
\centering
\subfigure{\includegraphics[scale=0.35]{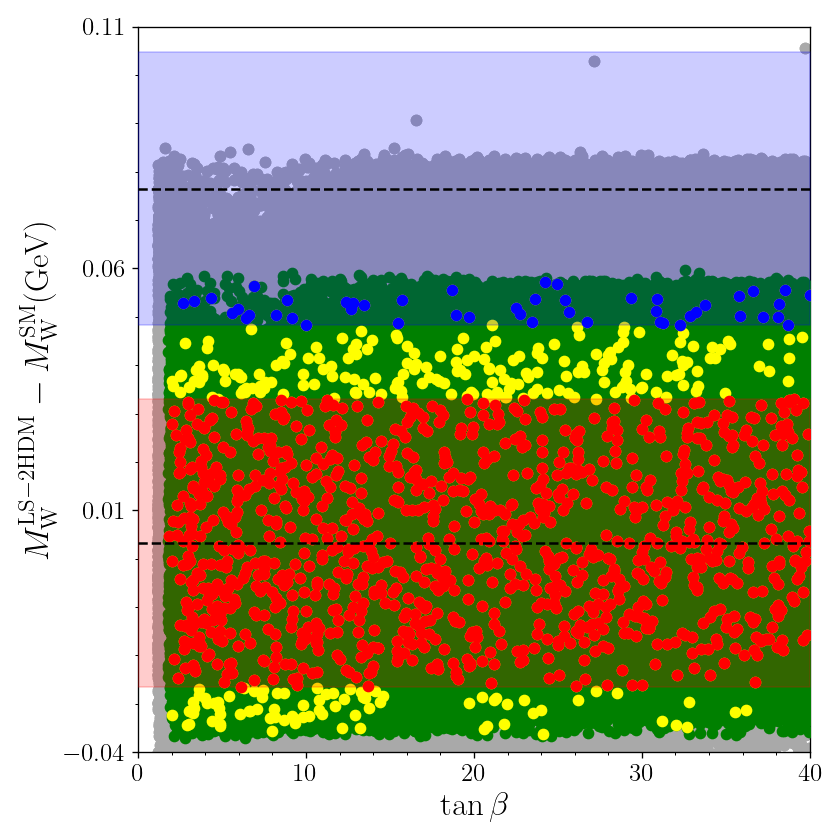}}%
\subfigure{\includegraphics[scale=0.35]{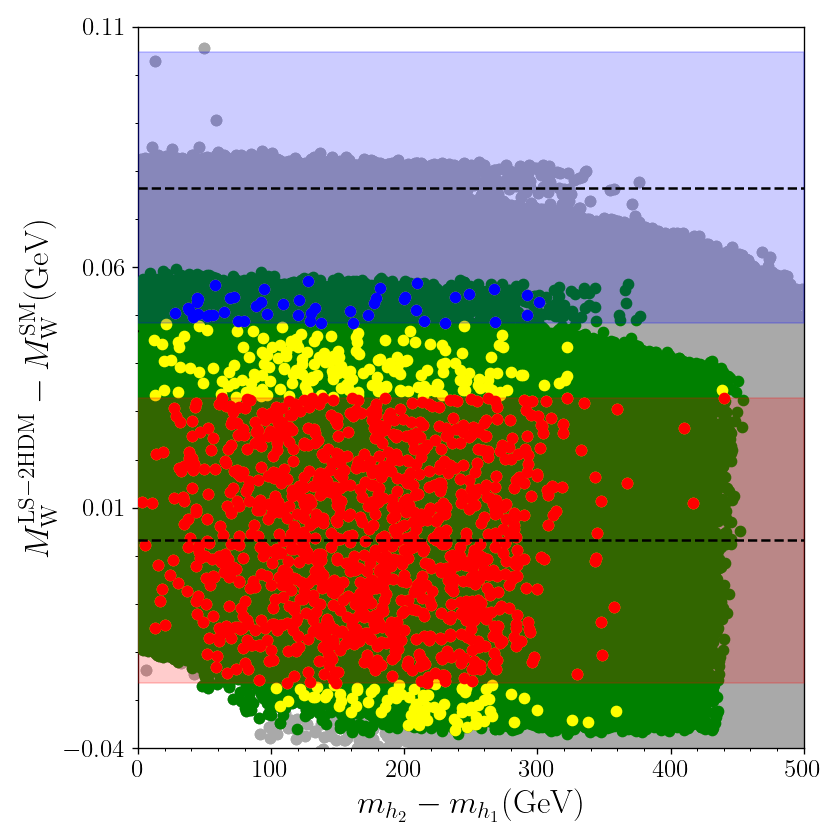}}\\
\subfigure{\includegraphics[scale=0.35]{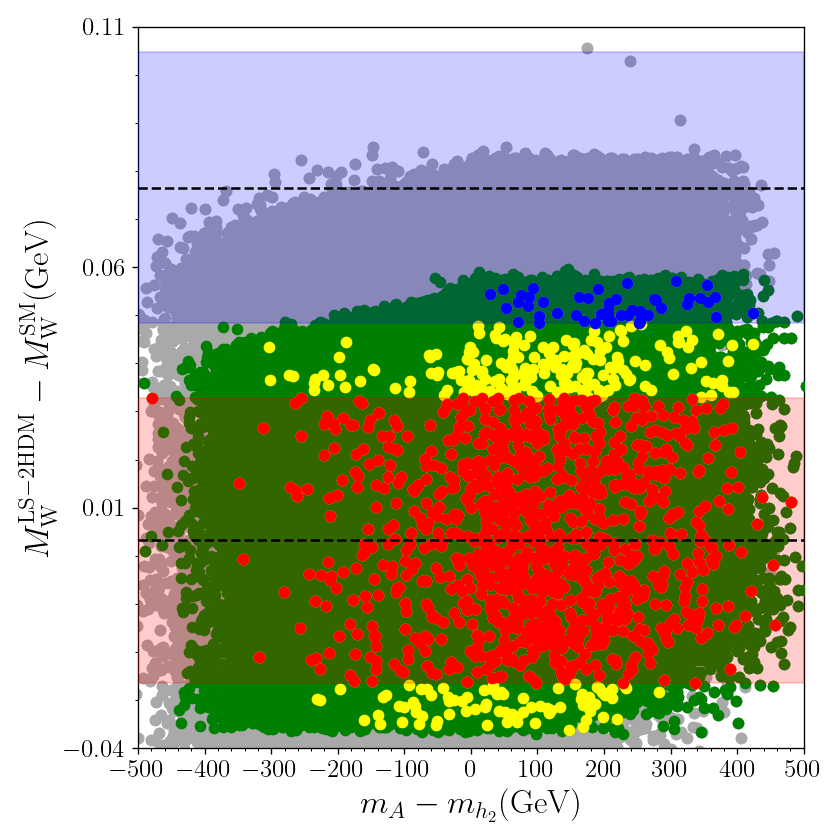}}%
\subfigure{\includegraphics[scale=0.35]{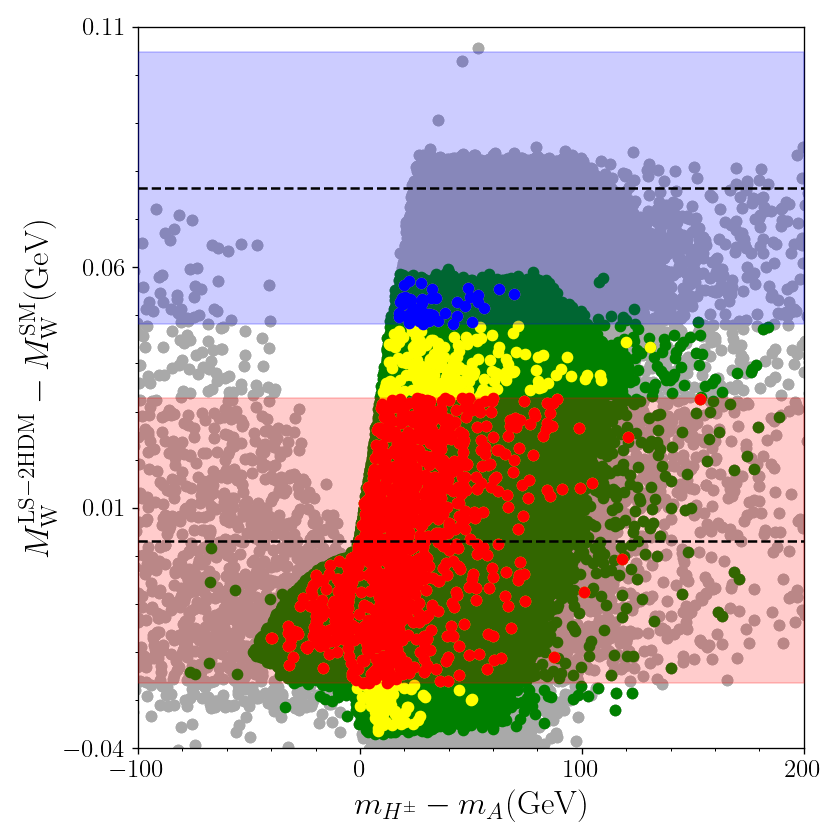}}
\caption{Solutions satisfying conditions G1, G2, G3, G4 and G5 plotted in the $\Delta M_W$ versus $\tan\beta$ (top-left), $m_{h_2} - m_{h_1}$ (top-right),  $m_{A} - m_{h_2}$ (bottom-left), and $m_{H^\pm} - m_{A}$ (bottom-right). Color coding of the points is the same as Fig. \ref{fig:Hmasses_1}. The blue region is the $3\sigma$ uncertainty range of the $M_W$ value measured by the CDF experiment and the red region is the $3\sigma$ uncertainty range of the $M_W$ value measured by the CMS experiment; the dashed lines indicate the corresponding published central values for the two measurements.}
\label{fig:W_mass_1}
\end{figure}

As seen from the equation which calculate the oblique parameters, the mass differences among the Higgs bosons and their mixing can be restricted directly. Their impacts on the mass differences are shown in Fig. \ref{fig:obliques} in correlation with the mass differences among thr Higgs boson. The bottom-right panel also displays possible correlation between $S$ and $T$ parameters. The color coding is the same as in Fig. \ref{fig:Hmasses_1}. The vertical and horizontal dashed lines indicate the solutions with $S,T,U = 0$ as referred to SM. As seen from the plots the solutions satisfying the constraints G1, G2 and G3 altogether (yellow points) yield the mass difference between the charged and CP-odd Higgs bosons between about -40 and 150 GeV. There is a slight increase in $T$ and $S$ parameters with increasing mass difference between these Higgs bosons. The correlation of $U$ parameter is seen sharper than the other oblique parameters. The CDF-compatible solutions for $W-$boson mass (blue points) are realized for large $T-$ parameter ($\sim 0.1-0.13$), while they are accumulated in both positive and negative neighborhood of the SM values (horizontal dashed line). We also display the $T$ and $S$ parameters with respect to each other to explore if there is any possible numerical correlation between them. Our scatter plots do not reveal any specific interval for these parameters with respect to each other. This is because our scans have several free parameters which can help fitting these parameters in any region consistently. On the other hand, if we pick up a point (say from red points), for such a point increasing $S$ values lead to an increase in $T-$parameter.

The restrictions on the mass differences from the oblique parameters also constrain the deviation in $W-$boson mass realized in our analyses. Fig. \ref{fig:W_mass_1} display the favored mass differences among the Higgs bosons by the desired deviation in $W-$boson mass compatible with the oblique parameters. The color coding is the same as in Fig. \ref{fig:Hmasses_1}. The horizontal dashed line at around zero indicates the CMS result, while the upper horizontal dashed line represents the CDF result. As seen from the top-left plane, the blue and red points do not prefer any specific value for $\tan\beta$. On the other hand, they restrict the mass difference between two CP-even Higgs bosons (top-right plane) at about 300 GeV by CDF (blue points), and at about 440 by CMS (red points) from upper. The mass difference favoured by blue ad red points can be as large as about 480 GeV for mass difference $m_{A}-m_{h_{2}}$. The main impact from the $W-$boson mass together with the consistent oblique parameters is seen in the mass difference between the charged and CP-odd Higgs bosons as shown in the bottom-right plane. To accommodate the desired deviation in $W-$boson mass requires $10 \lesssim m_{H^{\pm}}-m_{A} \lesssim 75 $ GeV by CDF and $-40 \lesssim m_{H^{\pm}}-m_{A} \lesssim 160 $ GeV by CMS.

Graphical representations reveal that the LS-2HDM scenario can generate a range of deviations in the mass of $W$ from small to large values, depending on the choice of appropriate parameters. Notably, one observes that the value $\Delta M_W$ exhibits variation,  encompassing both the Standard Model expectation and the experimental deviations in the CDF measurements. The findings confirm that the LS-2HDM model has the capability to maintain the $W-$boson mass in approximate ranges of the Standard Model, as upheld by a zero or insignificant $\Delta M_W$; e.g., $\Delta M_W \approx 0.003$ GeV, in alignment with CMS findings. When all other theoretical and experimental constraints are applied, except for the oblique-parameter bounds $(S,T,U)$, the LS-2HDM can yield solutions compatible with the CDF $W$-boson mass within about $1\sigma$; for example, one can obtain $\Delta M_W \sim 0.07$ GeV. However, once the current constraints on $S$, $T$, and $U$ are included, the deviation from the CDF value cannot be reduced below about $2\sigma$.

\subsection{Further Analysis}

\begin{figure}[t!]
\centering
\subfigure{\includegraphics[scale=0.35]{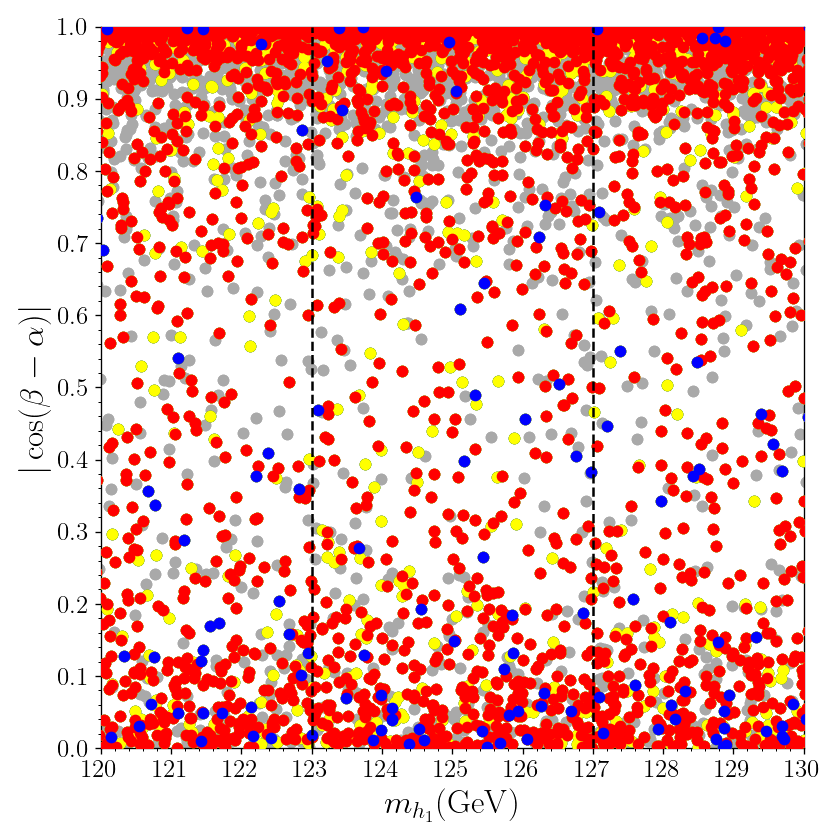}}%
\subfigure{\includegraphics[scale=0.35]{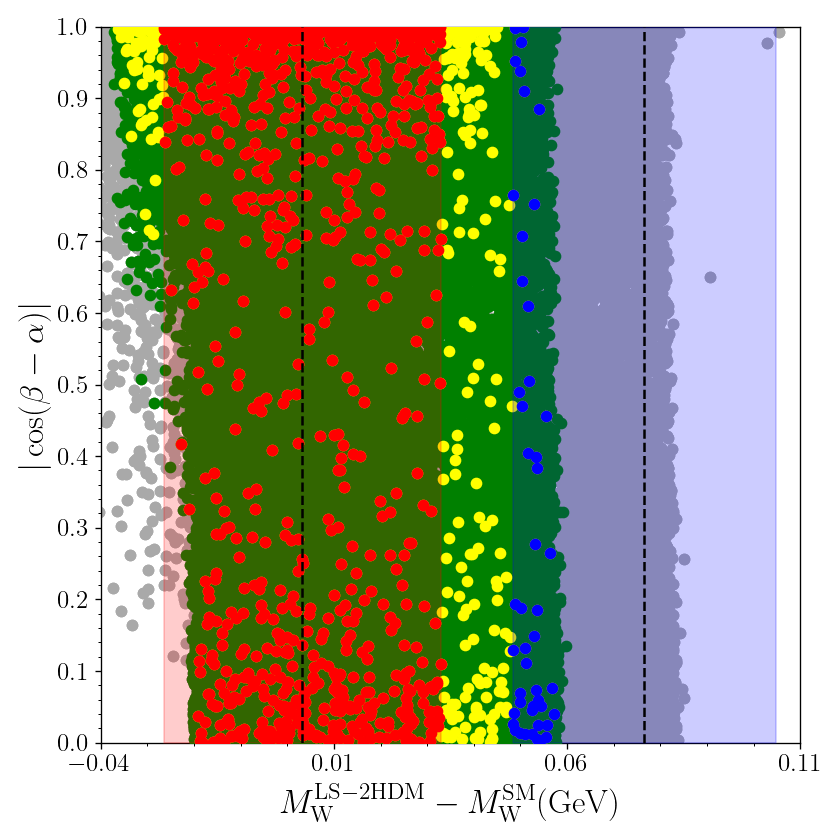}}\\
\subfigure{\includegraphics[scale=0.35]{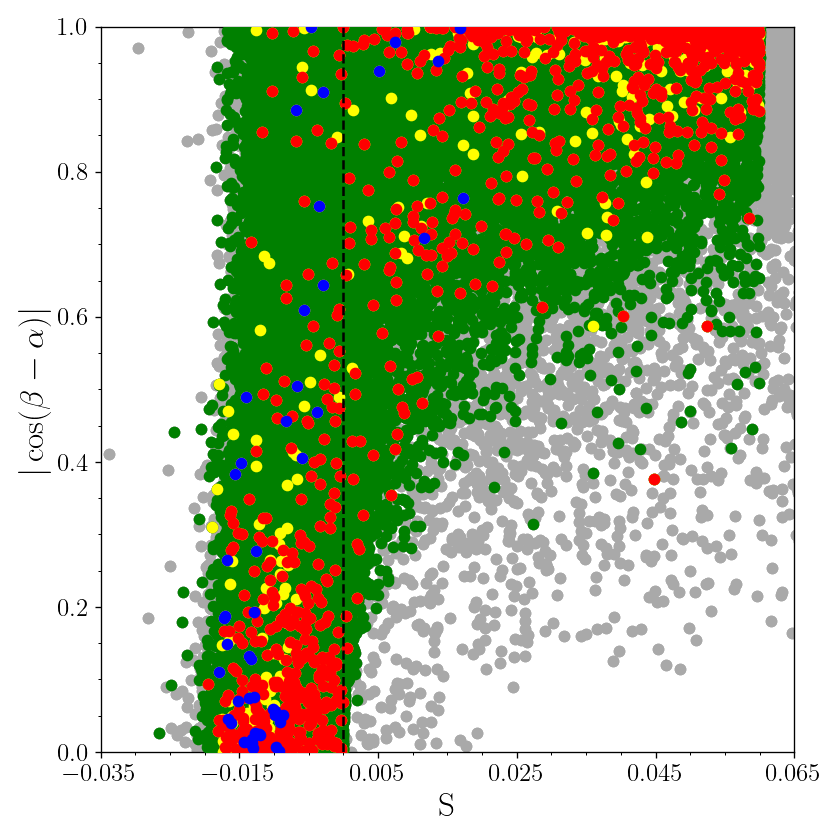}}%
\subfigure{\includegraphics[scale=0.35]{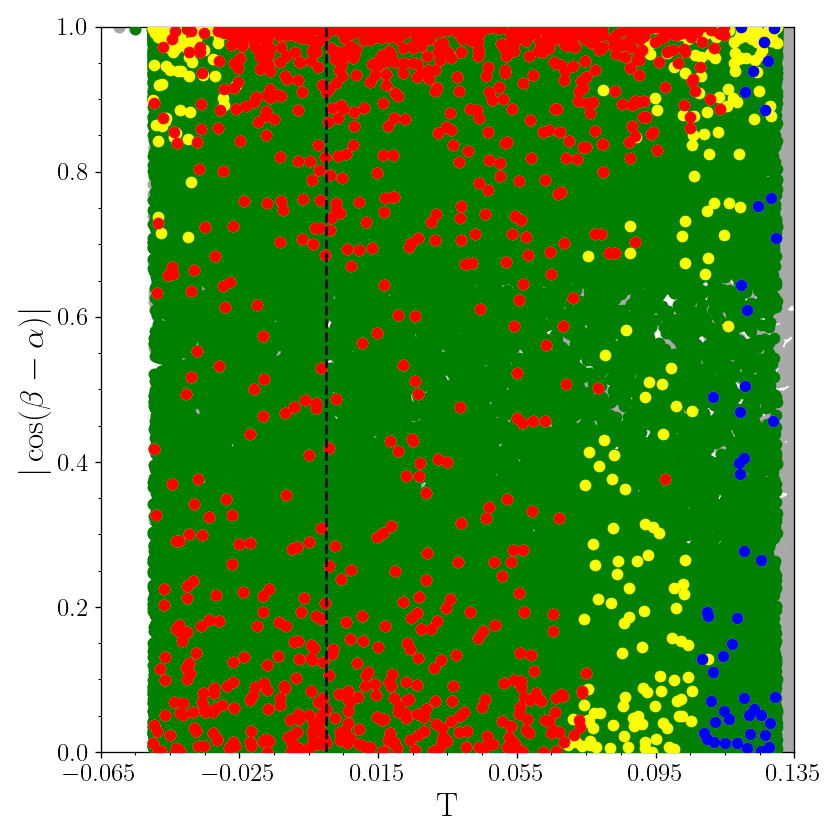}}
\caption{The plots for $\cos(\beta - \alpha)$ in correlation with the SM-like Higgs boson mass, deviation in $W-$boson mass and the oblique parameters. The color coding is the same as in Fig. \ref{fig:Hmasses_1}. The dashed lines in the top-left panel indicate the experimental bounds on the SM-like Higgs boson mass, while those in the top-right panel show the CMS and CDF measurements on $W-$boson mass from bottom to top. The dashed lines in the bottom planes show the solutions with $S,T = 0$ referring to SM.}
\label{fig:cosba}
\end{figure}

In previous sections, we have explored the deviations in $W-$boson mass confronting with several experimental and theoretical results, and we find that the LS-2HDM accommodates the CMS value and approaches the CDF value only up to about $2\sigma$ once the current $(S,T,U)$, SM-like Higgs, and LFU constraints are imposed. Despite a large variety of the constraint sets in our analyses, these solutions should be subjected in further analyses. Especially the properties of the SM-like Higgs boson is of a special importance in these further analyses. We have so far considered only a consistent mass range for the SM-like Higgs boson. A first step to investigate the features of the SM-like Higgs boson within the LS-2HDM framework is to consider the mixing among the doublets which is parametrized with $\cos(\beta - \alpha)$. We display the possible $\cos(\beta - \alpha)$ values in our analyses in Fig. \ref{fig:cosba} in correlation with the SM-like Higgs boson mass, deviation in $W-$boson mass and the oblique parameters. The color coding is the same as in Fig. \ref{fig:Hmasses_1}. The dashed lines in the top-left panel indicate the experimental bounds on the SM-like Higgs boson mass, while those in the top-right panel show the CMS and CDF measurements on $W-$boson mass from bottom to top. The dashed lines in the bottom planes show the solutions with $S,T = 0$ referring to SM. As seen from the top-left panel, a consistent mass for the SM-like Higgs boson can be realized in any value of $\cos(\beta - \alpha)$. The boundaries for this parameter reveals the SM-like Higgs boson formation. For instance, when $\cos(\beta - \alpha) \simeq 1$, the SM-like Higgs boson is mostly formed by the fields in $\Phi_{1}$, while those in $\Phi_{2}$ mostly form the SM-like Higgs boson when $\cos(\beta - \alpha) \sim 0$. The intermediate values of $\cos(\beta - \alpha)$ correspond to the mixing in the SM-like Higgs boson in which all the scalar fields take part actively. As seen from the top-right panel, one can accommodate the desired deviation in $W-$boson mass in all the range of $\cos(\beta - \alpha)$, and the bottom panels show these solutions can be consistent with the oblique parameters as well.

Even though we subject our solutions to the experimental and theoretical constraints strictly, still it does not mean that the resultant blue and red solutions are fully consistent. The first theoretical constraint on composition of the SM-like Higgs boson arises from the perturbativity limit, especially from the effective yukawa coupling with the top quark. The heavy mass for the top quark can be accommodated with a large coupling. In this context, a consistent SM-like Higgs boson should be formed by the doublet which directly interact with the top quark. In our set up, the suitable candidates for the SM-like Higgs boson are those resided in $\Phi_{2}$. Apart from the perturbativity limit, if the SM-like Higgs boson is formed by the fields in $\Phi_{1}$, then it would contradict with some other experimental results which are examining the effective coupling between the SM-like Higgs boson and gauge bosons as well as with the SM fermions. The most comprehensive way to check the consistency in the SM-like Higgs boson further from its mass is to run the HiggsTools package \cite{Bahl:2022igd}. This package checks mixing of the SM-like Higgs boson, its couplings to the SM particles and also confront the results with a large set of distinct experimental results in terms of its production and decay modes. Based on the discussion about the effective coupling between the SM-like Higgs boson and top quark, one can expect that the solutions forming the SM-like Higgs boson with the fields in $\Phi_{2}$ can survive after the HiggsTools run, which correspond to $cos(\beta - \alpha) \simeq 0$. 

    \begin{table}[h!] 
     \centering 
     \begin{tabular}{|l|c|c|c|c|c|c|}  \hline  
     \textbf{Parameters} & \textbf{Point 1} & \textbf{Point 2} & \textbf{Point 3} & \textbf{Point 4} & \textbf{Point 5} & \textbf{Point 6} \\ \hline  

    $M_{\rm W}^{\rm LS-2HDM}$ &  80.4142 & 80.4032 & 80.3786 & 80.3704 & 80.3540 & 80.3684 \\ \hline  

    $M_{\rm W}^{\rm LS-2HDM} - M_{\rm W}^{\rm CMS} $ &  0.054  \textcolor{red}{($5.87\sigma$)} &   0.0429 \textcolor{red}{($4.67\sigma$)}& 0.0184  \textcolor{red}{($1.86\sigma$)}& 0.0102 \textcolor{red}{($1.03\sigma$)} & -0.0062 \textcolor{red}{($0.63\sigma$)} & 0.0082 \textcolor{red}{($0.83\sigma$)}\\ \hline  

    $M_{\rm W}^{\rm LS-2HDM} - M_{\rm W}^{\rm CDF} $ & -0.0193 \textcolor{red}{($2.05\sigma$)}  &  -0.0303 \textcolor{red}{($3.22\sigma$)} & -0.0549 \textcolor{red}{($5.84 \sigma$)}  & -0.0631 \textcolor{red}{($6.71 \sigma$)}  & -0.0795 \textcolor{red}{($8.46\sigma$)}  & -0.0651 \textcolor{red}{($6.92\sigma$)}  \\ \hline  

    $m_{h_1}$ & 124.139 & 124.228 & 125.7700 & 124.1860 & 124.7140 & 125.9400 \\ \hline  

    $m_{h_2}$ & 252.081 & 246.601 & 144.5190 & 244.6180 & 233.2980 & 160.6120 \\ \hline  

    $m_{A}$ & 560.419 & 618.417 & 435.2390 & 536.0650 & 582.1850 & 532.8540 \\ \hline  

    $m_{H^\pm}$ & 582.554 &  633.576 & 443.2380 & 540.5340 & 579.4030 & 535.4250 \\ \hline  

    $m_{h_2} - m_{h_1}$ & 127.9420 & 122.3730 & 18.7490 & 120.4320 & 108.5840 & 34.6720 \\ \hline  

    $m_{H^\pm} - m_{A}$ & 22.1349 & 15.1589 & 7.9990  & 4.4690 & -2.7820 & 2.5710 \\ \hline  

    $m_{A} - m_{h_2}$ & 308.3379 & 371.8160 & 290.7200 & 291.4470 & 348.8870 & 372.2420 \\ \hline  

    $m_{H^\pm} - m_{h_1}$ & 458.4149 & 509.3480 & 317.4680 & 416.3480 & 454.6890 & 409.4850 \\ \hline  

    $\tan\beta$ & 24.2350 & 15.36 & 32.4200 & 20.4000 & 17.8040 & 23.7880 \\ \hline  

    $|\cos(\beta - \alpha)|$ & 0.0707 & 0.0446 & 0.0285 & 0.0464 & 0.0598 & 0.0442 \\ \hline  

    $\lambda_1$ & 2.7537 & 1.1050 & 0.4953 & 1.0403 & 2.2846 & 0.7583 \\ \hline  

    $\lambda_2$ & 0.1264 & 0.1257 & 0.1304 & 0.1263 & 0.1273 & 0.1307 \\ \hline  

    $\lambda_3$ & 10.12 & 12.303 & 6.1309 & 8.6155 & 10.2310 & 9.0426 \\ \hline  

    $\lambda_4$ & -4.9746 & -5.9280 &-3.0126 & -3.9144 & -4.5980 & -4.3505 \\ \hline  

    $\lambda_5$ & -4.1400 & -5.3019 & -2.7808  & -3.7557 & -4.7046 & -4.2599 \\ \hline  

    $m_{\rm 3}^2$ & -2598.6000 & -3955.5000 & -642.4600 & -2918.4000 & -3008.1000 & -1077.6000 \\ \hline  

    $B_s \rightarrow \mu^+ \mu^-$ & $3.072\times 10^{-9}$ & $3.073\times 10^{-9}$ & $3.072\times 10^{-9}$ & $3.073\times 10^{-9}$ & $3.073\times 10^{-9}$ & $3.072\times 10^{-9}$ \\ \hline  

    $B_s \rightarrow X_s \gamma $ & $3.149\times 10^{-4}$ & $3.148\times 10^{-4}$ & $3.149\times 10^{-4}$ & $3.149\times 10^{-4}$ & $3.148\times 10^{-4}$ & $3.149\times 10^{-4}$ \\ \hline  

    $ \delta_{\mu\mu} $ & $-3.920\times 10^{-5}$  & $-1.954\times 10^{-5}$ & $-3.718\times 10^{-5}$ & $-2.037\times 10^{-5}$ & $-1.890\times 10^{-5}$ & $-3.162\times 10^{-5}$ \\ \hline  

    $ \delta_{\tau\tau} $ & $-1.108\times 10^{-3}$& $-5.526\times 10^{-3}$ & $-1.051\times 10^{-2}$  & $-5.760\times 10^{-3}$  & $-5.344\times 10^{-3}$ & $-8.941\times 10^{-3}$ \\ \hline  

    $ g_\tau / g_\mu $ & 0.9996 &0.9998 & 0.9992 &0.9997  & 0.9998 & 0.9996 \\ \hline  

    $ g_\tau / g_e $ & 0.9996 &  0.9998 & 0.9992 & 0.9997  & 0.9998 & 0.9995 \\ \hline  

    $ g_\mu / g_e $ & 1.0000 & 1.0000 & 0.9999 & 1.0000 & 1.0000 & 1.0000 \\ \hline  

    $ S $ & -0.0162 & -0.0165 &  -0.0181 & -0.0145 & -0.0151 & -0.0180 \\ \hline  

    $ T $ & 0.128 & 0.1013 & 0.0405 & 0.0231 & -0.0169 & 0.0161 \\ \hline  

    $ U $ & 0.0006 & 0.0004 & 0.0004 & 0.0002 & -0.0000 & 0.0001 \\ \hline  

    $ \Delta a_\mu $ & $2.722\times 10^{-12}$  & $1.206\times 10^{-12}$ &$1.527\times 10^{-11}$ & $2.026\times 10^{-12}$ & $1.790\times 10^{-12}$ & $6.926\times 10^{-12}$ \\ \hline  
    \end{tabular} 
     \caption{Comparison of parameters across all points. All points passed from HiggsTools } 
     \label{tab:mw8035}
     \end{table}

Indeed, the impact from the consistency of the SM-like Higgs boson is further than this expectation such that only 10 points could survive out of about 1000 (red and blue points in total in the previous plots).  We exemplify these surviving solutions with 6 benchmark points in Table \ref{tab:mw8035}.
Points 1 and 2 exemplify the solutions which lead to $W-$boson mass compatible with the CDF results. These solutions accommodate the CP-odd and Charged Higgs boson masses at about 500-600 GeV. The interesting observation for these points that the CDF results rather imply relatively large mass difference between these two Higgs bosons at about 15-20 GeV. The heavy CP-even Higgs boson weigh around 250 GeV in these solutions. The consistent SM-like Higgs boson mass is realized at about 124 GeV, which is consistent with the observations within the theoretical uncertainties in Higgs boson mass calculation. Point 1 among these solutions also implies $1\sigma$ deviation from LFU in $\delta_{\tau\tau}$, while Point 2 leads to a $2\sigma$ deviation. Points 3,4,5, and 6 in Table \ref{tab:mw8035} represent the solutions compatible with the CMS measurements. These solutions can include the SM-like Higgs boson mass slightly heavier ($\sim 125$ GeV). As observed from these points, the charged and CP-odd Higgs bosons are realized nearly degenerate in mass. These solutions can be realized with relatively $\tan\beta$ as shown in Point 3. Despite relatively $\tan\beta$ ($\sim 32$), the muon anomalous magnetic moment ($\Delta a_{\mu}$) lies in the ranges consistent with the recent measurements \cite{Muong-2:2025xyk,Aliberti:2025beg}. Point 3 can lead to compatible new physics contributions to $\Delta a_{\mu}$, while predictions in Points 4, 5, 6 more or less coincide with the SM prediction.

Before concluding, we also present a possible limitations in the parameter space of LS-2HDM in Table \ref{tab:model_parameter_space}. Even though we generate the solutions by varying the couplings, we express the ranges for masses and mass differences to provide a comparable values with the other studies. Note that the CDF and CMS analyses reveal different ranges for the deviation in $W-$boson mass and these ranges do not overlap. Therefore we display two sets of ranges. The first two columns display the minimum and maximum ranges when the CDF results are imposed, while the last two columns show the limits for the case of CSM results.

        \begin{table}[h!]
        \centering
    \renewcommand{\arraystretch}{1.5} 
    \setlength{\tabcolsep}{16pt}
        \begin{tabular}{|c|c|c|c|c|c|}
    \hline
    Parameter &Min. Val. (CDF) & Max. Val. (CDF)  & Min. Val. (CMS) & Max. Val. (CMS) \\
    \hline
    $m_{h_1}$ & $124.139$ & $124.228$ & $123.668$ & $125.940$ \\
    \hline
    $m_{h_2}$ & $246.601$ & $252.081$ & $144.519$ & $413.886$ \\
    \hline
    $m_{A}$ & $560.419$ & $618.417$ &$435.239$ & $671.085$\\
    \hline
    $m_{H^\pm}$ & $582.554$ & $633.576$ & $443.238$  & $685.670$\\
    \hline
    $\tan\beta$ & $15.36$ & $24.235$ & $17.804$  & $38.741$\\
    \hline
    $|\cos(\beta - \alpha)|$ & $0.039$ & $0.071$ & $0.023$  & $0.060$\\
    \hline        
    $m_3^2$ & $-3955.5$ & $-2598.6$ & $-4926.5$  & $-642.46$\\
    \hline
    $m_{h_2} - m_{h_1}$ & $122.373$& $127.942$ & $18.749$ & $290.218$ \\
    \hline
    $m_{A} - m_{h_2}$ & $308.338$& $371.816$ & $253.406$ & $372.242$ \\
    \hline
    $m_{H^\pm} - m_{A}$ & $15.159$& $22.135$ & $-11.121$ & $15.588$ \\
    \hline
        \end{tabular}
        \caption{The constraints on the model parameters which satisfy all four aforementioned group of limitations and HiggsTools. All masses are given in GeV unit.}
        \label{tab:model_parameter_space}
    \end{table}
\section{Conclusions}

In this work, the LS-2HDM parameter space is investigated with the aim of obtaining a parameter space where the $W-$boson mass measured by CDF and CMS experiment is confronted within the framework of LS-2HDM. To this end, both theoretical and experimental constraints are applied to the parameter space. Consequently, only a limited range of parameter values remains viable within the LS-2HDM framework, most of which are found to be compatible with the reported measurements of LFU tau-lepton and $Z-$boson decays. Among these, the electroweak precision constraints encoded in the oblique parameters $(S,T,U)$ play the crucial role. Enforcing the current global fits excludes solutions that would otherwise realize the CDF $1\sigma$ region, whereas the charged-Higgs mass and rare $B$-decay bounds have a comparatively weak effect. Furthermore, assuming $h_1$ to be the SM-like Higgs boson reduced the number of solutions considerably. In our scans we required consistency with the current $(S,T,U)$, SM-like Higgs, and LFU constraints, and then tested compatibility with the CDF and CMS $M_W$ determinations. It is clear that, within the feasible solutions, the LS-2HDM easily predicts the CMS measurement, while the predicted $W$-boson mass can approach the CDF value only to about $2\sigma$ once all constraints are imposed. As a result of this analysis, the available parameter space of LS-2HDM is summarized in Table \ref{tab:model_parameter_space}.  Within these ranges, LS-2HDM can be effectively utilized for making predictions. 

It is important to note that imposing the ($M_{\rm W}^{\rm LS-2HDM} - M_{\rm W}^{\rm CMS/CDF}$) condition directly on the masses may not always be convenient for determining restrictions on parameters. Using additional parameters related to masses, such as the mass differences employed in this analysis, results in stronger limits. We observe that the current constraints on the oblique parameters can directly bound the mass differences such that the consistent solutions can accommodate CDF measurements for the W-boson mass only up to about $2\sigma$, and it further tightened by requiring $h_1$ to be SM-like and by the LFU constraints. This discrepancy between $h_1$ being SM-like and $M_W^{\rm CDF}$ might be a sign of an inconsistency that requires further investigations. 

To ensure a comprehensive analysis, six solutions were selected, with two estimating $M_{\rm W}^{\rm CDF}$  within $2\sigma$ and remaining four predicting  $M_{\rm W}^{\rm CMS}$  within $1\sigma$. These selected solutions were used as benchmarks, and their predictions were provided. To comprehensively conclude this analysis, all solutions satisfying the conditions outlined in Table \ref{tab:model_parameter_space} were tested using the HiggsTools package, which incorporates the most recent constraints. Particularly with the addition of the new value of the $W-$boson mass derived by CMS in 2024 in the model, it can be seen that the LS-2HDM has a parameter space that accommodates the CMS measurement and approaches the CDF value about $2\sigma$, but does not resolve the CDF and CMS tension within a single framework.

\begin{acknowledgements}
The authors would like to extend their sincere gratitude to Cem Salih Ün for the invaluable support and expertise provided throughout this study. Part of the computational work reported in this paper were conducted at the High Performance and Grid Computing Center (TRUBA Resources) of the National Academic Network and Information Center (ULAKBIM) of TUBITAK. Additionally, the authors acknowledge partial support from the Turkish Energy, Nuclear and Mineral Research Agency under project number 2022TENMAK(CERN)A5.H3.F2-02.\vskip0.5cm

\end{acknowledgements}
\section*{Data Availability Statement} 

Data sets generated during the current study are available from the corresponding author on reasonable request.

\bibliographystyle{apsrev4-2}   
\end{document}